# How assigned AI use before class shapes active student engagement in class

Dan J. Wang[1,*]
Neelam Modi Jain[1]
Vanessa Burbano[1]
Jorge Guzman[1]
Daniel Keum[1]
Soomi Kim[1]
Bruce Kogut[1]
Nataliya Wright[1]

*Author Affiliations:*
[1] Management Division, Columbia Business School, Columbia University, New York, NY 10027, USA.

** Corresponding Author:*
Dan J. Wang, Lambert Family Professor of Social Enterprise in the Faculty of Business, Management Division, Columbia Business School, Columbia University, New York, NY 10027, USA
Email: djw2104@gsb.columbia.edu

**Abstract**

AI learning tools are rapidly entering classrooms, but evidence about whether they help students learn is mixed and rests mostly on test scores. Comparatively less research addresses whether the use of AI changes students' live learning behaviors in class. Here, we report the results of a preregistered field experiment with 759 MBA students enrolled in ten sections of a course, in which each student was randomly assigned two of ten class sessions to prepare for with a purpose-built voice-based AI discussion partner. After two uses of the AI discussion partner, students made about 31% more voluntary contributions in each later class session. Students who used the AI discussion partner more also reported greater comfort speaking up and greater perceived learning, but not greater focus or motivation. These findings suggest that repeated practice with a voice-based AI partner can meaningfully increase students' engagement in class discussion, enhancing a critical intermediate learning outcome.

## Introduction

Can using AI outside of the classroom stimulate learning processes inside the classroom? The use of AI has pervaded educational settings as students, instructors, and school administrators wade through an ever-shifting landscape of findings, practices, and policies to decide how best to make use of AI to enhance learning outcomes. Not limited to the domain of learning science, researchers from across the social sciences have taken a variety of approaches – from randomized controlled trials and large-scale observational analysis to in-depth case studies – to grasp the implications of widespread AI use in classroom environments. Yet, the findings from this work point in different directions, with evidence that permitting, encouraging, or requiring AI can be both beneficial and detrimental to student achievement.

Whereas many of these studies target *assessment-based outcomes* in examining the effect of student AI use, less scholarship has addressed how using AI might affect intermediate learning behaviors that are just as critical to students' experiences. We analyze how the introduction of a novel voice-based AI tool intended to help students absorb material in preparation for class affects students' propensity to participate in class discussion. We report the results of a pre-registered field experiment among 759 MBA students across 10 sections of a required course, in which in-class discussion was an important element of the pedagogical approach. In the study, each student was randomly assigned to two out of 10 sessions, for which they were permitted to use a voice-based AI discussion partner tool, called CAiSEY, to prepare. We find that after using CAiSEY twice, a student's contribution count in each subsequent class session was 31% higher than the average count of contributions per student per session. Furthermore, after using CAiSEY more throughout the semester, students reported an increase in their comfort with class participation and in their own learning of class material.

Our study shifts the focus of research on AI and education away from assessment-based outcomes to a learning behavior that connects with real-world achievement. Participating in class demonstrates confidence, mastery of knowledge, and self-efficacy. Speaking up among one's peers is itself both a performance for which one has to prepare as well as a form of practice for real-world settings in which many decisions are made through live, synchronous, oral-based communication. Despite the widespread acknowledgment that spoken rhetorical ability is a core lifelong skill that students gain through class participation, surprisingly few studies have examined how pedagogical interventions might motivate students to take greater initiative to participate in class, much less the role that AI tools might play.

We situate our study within the broader literature on technology use in classroom settings, which has produced mixed results about how generative AI contributes to student learning. Because the development of generative AI tools for education is nascent, research that examines its consequences represents a diversity of deployment strategies, classroom environments, and assessment approaches. As a result, even recent meta-analyses of whether using AI boosts student testing outcomes have been questioned or even retracted (Deng et al. 2025). Therefore, research on whether encouraging or mandating AI use benefits student learning depends on a host of conditions that has not yet converged on a generalizable framework.

Research evidence that AI undermines learning tends to rely on the general idea that AI tempts students to engage in *cognitive offloading*, a behavior in which students become over-reliant on an AI tool to work through a learning task (Bastani et al. 2025, Lehmann, Cornelius, and Sting 2024, Gerlich 2025). The effect is akin to a student relying on a tutor, who is too willing to reveal the solution to a problem rather than forcing students to work through a learning barrier themselves. By contrast, other work has shown that when instructors can tailor the use of AI tools with course-specific pedagogical guardrails, student learning outcomes benefit from using AI (Kestin et al. 2025, De Simone et al. 2025). Furthermore, under such conditions, students also report that they feel more motivated and engaged in the classroom.

Reconciling these findings, however, remains a challenge because in many settings, there is limited visibility into the mechanisms that link student AI use to the learning outcomes used to benchmark the effectiveness of AI tools. Specifically, most studies center around standardized assessment, typically, a grade on an exam or a score on a homework assignment, as an indicator of learning success (Bastani et al. 2025, Kestin et al. 2025, Wecks et al. 2024, Hausman, Rigbi, and Weisburd 2025). Missing from this work is insight into the intermediate processes that might also be triggered by AI use and in turn, explain the variation in assessment-based outcomes. Moreover, some intermediate learning processes might themselves be overlooked as vital indicators of learning and skill development that are directly relevant to predicting achievement in other domains (Fredricks, Blumenfeld, and Paris 2004, Heckman, Stixrud, and Urzua 2006, Deming 2017).

Our study prioritizes student participation in class discussion as an outcome for three reasons. First, class participation captures dimensions of skill development and learning, such as self-efficacy and critical thinking in addition to mastery of knowledge, that other assessment-based learning outcomes cannot. Second, class participation is itself a skill that generalizes to settings outside the classroom, predicting career mobility, leadership success, and other outcomes that are a function of rhetorical and social skill. Finally, class participation is an indicator of student motivation and engagement, which according to prior work, lead to greater student learning in class.

Although the importance of class participation as a generally desirable learning behavior is well established, empirical work on ways to mobilize more participation from students has mostly examined in-class tactics that instructors might deploy. Work on pre-class interventions, such as rehearsing with a classmate before class (Mundelsee and Jurkowski 2021), demonstrates some promising evidence, but such techniques are difficult to enforce with standardization and visibility. Moreover, work on using AI tools to help with speaking confidence has not examined classroom behavior, focusing instead on self-reported feelings and individual progress in communication skill (Ayedoun, Hayashi, and Seta 2019, Tai and Chen 2023, Tai 2024, Mukadam, Suresh, and Jacobs 2025, Tyrrell et al. 2025).

Our field experiment directly addresses these shortcomings. To our knowledge, our study is the first to incorporate a randomized controlled trial to examine how a generative AI tool deployed at scale might affect students' in-class participation behavior. Unlike prior work, our study sample consists of students in a professional degree program, rather than a primary, secondary, or undergraduate classroom setting. In addition, we obtain data from directly observed in-class behavior and each student's self-reported engagement and learning throughout the term. Our approach therefore permits a causal interpretation of the effects on student behavior from a scalable generative AI-enabled classroom intervention.

## Motivation from Prior Literature

### *Class Participation as a Learning Outcome*

Stimulating student participation in classroom discussions has long been an essential component in an instructor's classroom toolkit. Student discussion in class comes in various forms, from a classwide debate to the Socratic method of teaching, wherein student contributions – either voluntary or elicited by 'cold calling' from the instructor – create an active learning experience. Research from learning science identifies two ways in which speaking up in the classroom leads to greater cognitive engagement, resulting in positive learning outcomes, such as the retention of knowledge and the application of knowledge in other settings. One, according to Chi, et al (2014), voluntary class participation affords students the opportunity to receive active feedback on their contributions that compel them to make "adjustments to their own mental model", an crucial indicator of critical learning activity. Two, speaking up also offers students the chance to explain what they have learned to their classmates and instructors. The act of explaining activates a generative learning process that deepens a student's understanding about an idea or topic they are studying (Kobayashi 2019).

In addition to contributing to learning outcomes, scholarship has also emphasized how contributing to a discussion through oral communication has implications for an individual's achievement in real-world settings. Prior to the advent of widely accessible generative AI tools, employer surveys consistently rated verbal communication (specifically, oral communication) as among the topmost valued skills in the workplace (AAC&U and Hart 2018). In the years since the emergence of generative AI – demarcated with the public release of ChatGPT in November 2022 – surveys continued to put verbal communication as one of the top three skills valued by recruiters (GMAC 2024, 2025). Furthermore, research from organizational behavior scholars has demonstrated that speaking up in the workplace predicts stronger job performance (Chamberlin, Newton, and LePine 2017, Whiting, Podsakoff, and Pierce 2008) and a greater likelihood of being promoted into leadership positions (MacLaren, et al 2020). Outside of the workplace, scholarship has shown that active learning through classroom discussion promotes greater civic engagement as well (Campbell 2008, Toney-Purta 2002, Kawashima-Ginsberg and Levine 2014).

Alongside the range of work describing the positive learning and achievement outcomes that result from engaging in more classroom discussion, there has been some work on interventions and tactics to stimulate discussion in the classroom. Within the set of studies that treat student participation in the classroom as an outcome variable, most work focuses on features of the classroom itself or techniques that can be deployed in the classroom by instructors. For example, studies have repeatedly demonstrated the effectiveness of cold-calling – an instructional technique whereby an instructor asks a student to contribute to a classwide conversation without the student having prior knowledge of being selected to do so – for motivating a student to engage in future voluntary participation (Dallimore, Hertestein, and Platt 2013, 2019, Edd, Brownell, and Wedenroth 2014, Waugh & Andrews 2020). Other studies have found that a more 'invitational' framing of discussion questions asked by instructors in the classroom can elicit greater participation (Frykenberg, et al 2023) while having each student's name publicly displayed on tent cards in front of them also induces greater participation (Cooper, et al 2017).

Empirical studies of extra-class interventions to encourage in-class participation are even more scarce. The study that most approaches the design of our field experiment comes from Ghosh, et al (2023), who randomly assign a subset of students in a physiology class to use a digital self-guided study tool – that was not based in generative AI – for a single session. The authors find that students who used the tool reported having a lower fear of negative evaluation by their peers, which they interpret as evidence for lowered anxiety about speaking up in class. The research design, however, did not capture actual behavioral outcomes in which participation could be observed. Besides Ghosh, et al (2023), however, we could identify no other study that investigated whether a student's contribution to class discussion could be causally linked to some relevant stimulus outside of the classroom.

*AI Tutors and Communication Skills as Learning Outcomes*

A core motivation for our study comes from the promise that AI-based tools can raise students' academic achievement. Even before generative AI tools, educational software producers had long touted adaptive approaches to guide student learning by simulating natural language interactions and feedback. The most well-known example of an approach that has scaled commercially is arguably Khan Academy, which has reached over 200 million student users through a variety of digital tutor-led course experiences. Even before Khan Academy, however, tools such as AutoTutor (Graesser, et al 2004, D'Mello, et al 2011), ITSPOKE (Litman, et al 2006), and other computer mediated tutoring programs enabled students to receive personalized and dynamic feedback on topics such as math and science in a learning session that simulated what it would be like learning from a human tutor. When subjected to randomized controlled trials, their usage has generally led to positive learning effects, especially for assessment-based outcomes such as math tests in K-12 environments, that are close to if not on par with those of human tutors.

Since the advent of generative AI, an emerging body of work has focused on how the use of general purpose chatbots, such as ChatGPT, might enhance individuals' communication skills more generally. For example, a recent study by Mukadam, et al (2025) followed 27 medical students who each used Open AI's GPT-4o voice mode to practice interacting with a patient in various scenarios. Interviews with the students following their experiences revealed that they largely felt more confident and found the spoken practice useful. A similar study by Yamamoto, et al (2024) found that among a sample 35 medical school students who practiced patient interactions on a text-based ChatGPT interface scored, on average, higher on a spoken interview exam compared with students in the past who did not have generative AI tools available to them. Finally, Tai (2024) found that using a voice-based AI language tutor led a group of 92 English-as-a-Foreign-Language (EFL) students to report a higher willingness to speak in English compared with two groups who only had human language tutors. Additional data analysis from this work revealed that their greater willingness could be attributed to reporting greater communicative confidence and lower speaking anxiety; moreover, the group with the AI language tutor reported overall lower levels of anxiety in their tutor sessions than the group assigned to human language tutors.

Taken together, these findings suggest that AI interventions that involve 1:1 feedback can help students feel more confident and less anxious about communicating with others. Yet, none of their empirical approaches deploy a randomized treatment design, foreclosing a causal interpretation of their results. In addition, none of these studies link experience with AI-based tutors to observable communication outcomes as they rely on self-reports about willingness to speak. Zooming out, the vast majority of studies on digital or AI tutors – and human tutors – emphasize learning outcomes that reflect objective assessment, marginalizing the importance of in-class voluntary participation as an important learning behavior. These studies therefore invite further targeted investigations of the connection between pedagogically intentional AI use and in-class student learning behaviors.

*Main Hypotheses*

We argue that interaction with a knowledgeable voice-based AI discussion partner stimulates greater class participation by elevating a student's mastery of knowledge and lowering their anxiety about speaking up in the classroom. Our reasoning comes from a long tradition of studies in educational research that has identified the Fear of Negative Evaluation (FNE) as what often hinders student learning and engagement in the classroom. FNE itself refers to anxiety about being criticized or judged by others (Leary 1983, Watson and Friend 1969). Individuals with high FNE tend to avoid situations in which they might provoke a negative judgment about themselves by others through social comparison (Friend and Gilbert 1973). In education research, FNE manifests prominently as a factor associated with lower academic motivation and self-regulation, among other factors, which in turn leads to lower academic achievement and worse learning outcomes (Toprak and Yavuz 2026, Elliot and Church 1997, Bandura 1997, Wolters 2003).

A subset of work on FNE focuses on classroom participation as an outcome, surfacing a set of mechanisms that underlie the consistent negative relationship between FNE and voluntary contributions in class. One such mechanism is self-efficacy, or the belief in one's own abilities to achieve a goal (Bandura 1997). Higher self-efficacy predicts greater classroom participation because students have greater confidence that they will contribute ideas, views, and insights that will result in positive feedback from an instructor or their classmates. Those with higher self-efficacy tend to believe they have achieved a deeper understanding of some domain or skill. Therefore, having greater mastery of a domain of knowledge can raise a student's self-efficacy, which in turn manifests as a higher tendency to participate in classroom discussions. Indeed, Yue, Jia, and Wang (2022) show in a study of 568 undergraduate nursing students that the relationship between lower self-efficacy and lower rates of class participation was largely mediated by greater FNE.

A second mechanism relates to social anxiety. Independent of one's understanding of a subject, anxiety about sharing one's thoughts in public can also contribute to greater FNE. Aside from the content of speech, anxiety about public speaking can come from concern about being judged negatively for *how* one speaks based on features that are unrelated to one's understanding of a subject. Here, a large body of research has shown that rehearsal and speech training can reduce such anxiety-related barriers (Hunter, Westwick and Helta 2014, Dwyaer and Davidson 2012). Such interventions focus on speaking and communication as a skill rather than helping an individual gain greater mastery about a topic.

We assess whether voice-based AI tools like CAiSEY might address both sources of FNE. Specifically, we examine how student usage of such tools might be related to the propensity to participate in class discussions, a form of public speaking that demonstrates both communication skill and mastery of knowledge in an environment of perceived social evaluation. Although emerging, a cohesive body of evidence points to the promising effects that AI-based training can help alleviate the anxiety individuals have with public oral communication more generally. In the realm of foreign language learners, Tai and Chen (2023) find that the use of a Google AI Assistant among a group of Taiwanese eighth graders learning English led to greater communicative confidence and willingness to communicate. Mediating the effect was that students felt less anxious not only about speaking after practicing with the AI assistant, but also *while* practicing with the AI assistant. Similar results obtained in studies of foreign language learners by Tai (2024), Ayedoun, Hayashi, and Seta (2019). Mukadam, Suresh, and Jacobs (2025) apply the same approach in following 27 medical students, who report greater confidence in communicating with patients following training with a voice-based AI. Finally, Tyrell, et al (2025) conduct a randomized controlled trail across almost 400 medical students in the UK who find that self-rated communication skills improved with an AI patient simulator nearly as much as with a human patient simulator.

Taken together, these findings guide our hypotheses, which summarize our argument about how practice with a voice-based AI discussion partner leads to differences in (1) a student's behavioral outcomes in the classroom discussion (H1) and (2) self-reported outcomes about a student's own learning experience (H2a-d).

H1. As students use CAiSEY more, the volume of their voluntary in-class participation increases.

H2a. As students use CAiSEY more, they report greater comfort in participating in the classroom.

H2b. As students use CAiSEY more, they report greater focus in the class.

H2c. As students use CAiSEY more, they report greater motivation in the class.

H2d. As students use CAiSEY more, they report having a deeper learning experience in the class.

**Results**

*Assignment and Use of CAiSEY*

We conducted the experiment in ten sections of a required first-year MBA strategy course at a U.S. business school during Fall 2025. Each of the 759 randomized students was assigned to use CAiSEY before two of the ten eligible case discussions. Instructors and teaching assistants for each class were blind to the random assignment. The design of the study permitted a student to access and use CAiSEY for only the two case discussions to which they were randomly assigned, prohibiting access to CAiSEY for any other case discussions. The experiment therefore randomizes when, rather than whether, each student was assigned to the tool, so that students assigned earlier or later in the course serve as one another's comparison. Students assigned earlier and later did not differ in pre-course comfort participating or years of work experience (Supplementary Table S14, Panel A: joint tests, $p > 0.1$).

Students took up the tool readily. Among the 504 students whose two assigned sessions were observed, between 92.2% and 98.9% of assigned students in each session submitted at least one conversation with CAiSEY, and between 38.1% and 67.0% held a conversation of at least ten minutes, which we define as qualifying use (Figure 1; Supplementary Table S1).

We evaluate CAiSEY in two complementary ways. Our models of *assignment estimates* capture the effect of being assigned to CAiSEY for a case, regardless of whether the student used it (the intention-to-treat, or ITT, effect). Our models of *qualifying use estimates* capture the effect of a conversation of at least ten minutes. We obtain these by using random assignment to isolate use that was prompted by the assignment, which yields a local average treatment effect (LATE) for students whose use responded to assignment (details in Methods). We report both for class participation and for self-reported learning experiences.

*[insert Figure 1 around here]*

*Class Participation*

H1 predicts that using CAiSEY more increases voluntary class participation. We measure participation as the number of voluntary contributions a student made in a session, which excludes contributions elicited by an instructor's cold call. Each voluntary contribution in a class by a student was recorded by a live in-class teaching assistant and later reviewed by the class instructor. In the 4,787 student-sessions from 504 students used for the models that distinguish a student's first and second assigned opportunity, students made 0.73 voluntary contributions on average (SD = 0.83) and contributed at least once in 52.9% of sessions (Supplementary Table S2, Panels A and B).

CAiSEY prepares students before class, so its effects could appear in the discussion of the case it was used for, in the following class, or across the rest of the course. We therefore test H1 at three horizons: the same session as the assigned case (Same Session), the next observed session (Next Session), and all later sessions (All Future Sessions). Within each horizon, we separate each student's first and second assigned opportunity, the earlier and later of their two assigned sessions, to ask whether effects depend on repeated use (Figure 2; Supplementary Table S4). The coefficients on the first and second assigned opportunity are $\beta_1$ and $\beta_2$, respectively (Equation 1).

We find no evidence that using CAiSEY changes participation in the discussion of the case itself. Estimates for the first and second assigned opportunity are close to zero for qualifying use (Supplementary Table S4, Model 3: $\beta_1$ = −0.040, SE = 0.073, p > 0.1; $\beta_2$ = −0.036, SE = 0.062, p > 0.1) and for assignment (Model 2: $\beta_1$ = −0.019, SE = 0.036, p > 0.1; $\beta_2$ = −0.019, SE = 0.033, p > 0.1), as is the pooled assignment estimate (Model 1: β = −0.011, SE = 0.025, p > 0.1).

The effects of CAiSEY emerge in later sessions and depend on a second assigned opportunity. After the second assigned opportunity, students participated more in the next session, whether measured by assignment (Supplementary Table S4, Model 4: $\beta_2$ = 0.091, SE = 0.043, p < 0.05) or by qualifying use (Model 5: $\beta_2$ = 0.183, SE = 0.087, p < 0.05). They also participated more across all future sessions (Model 6: $\beta_2$ = 0.112, SE = 0.046, p < 0.05; Model 7: $\beta_2$ = 0.227, SE = 0.096, p < 0.05; Figure 2). Because the All Future Sessions model keeps the first assigned opportunity in effect after the second, $\beta_2$ in these models is the additional effect of a second use. The qualifying use estimate (Model 7) corresponds to about one additional voluntary contribution for every four to five sessions, or roughly 31% of the sample mean of 0.73 contributions per session (Supplementary Table S2, Panel A).

*[insert Figure 2 around here]*

A first assigned opportunity shows no such benefit. In the next session, participation was lower after qualifying use at the first assigned opportunity (Supplementary Table S4, Model 5: $\beta_1$ = −0.151, SE = 0.066, p < 0.05), and across all future sessions it was not detectably different from zero (Model 7: $\beta_1$ = −0.059, SE = 0.087, p > 0.1; Figure 2).

Figure 3 illustrates how the additional effect of a second assigned opportunity unfolds over the course. It compares predicted participation for a student assigned in sessions 3 and 5 with that of a student assigned in session 3 alone, so that the two differ only by the second assigned opportunity in session 5. Because every student received two assignments, the one opportunity schedule is a model-based benchmark rather than an observed group. The predicted difference is near zero in session 5 and positive in every later session, from 9% to 22% of predicted participation under assignment and from 20% to 49% under qualifying use (Figure 3; Supplementary Table S7). Summed over sessions 5 through 12, a second assigned opportunity corresponds to additional contributions per student (Supplementary Table S6, Model 1: Δ = 0.652, SE = 0.317, p < 0.05; Model 2: Δ = 1.333, SE = 0.664, p < 0.05).

*[insert Figure 3 around here]*

We next ask whether these effects require longer conversations, by redefining qualifying use as an interaction of at least four minutes, five minutes, and so on up to ten minutes (Figure 4; Supplementary Table S5). The effect of a second assigned opportunity is positive and statistically significant at every threshold. It ranges from $\beta_2$ = 0.099 to 0.183 in the next session (Supplementary Table S5, Panel B, Models 15 to 21; all p < 0.05) and from $\beta_2$ = 0.122 to 0.227 across all future sessions (Panel C, Models 22 to 28; all p < 0.05). The benefit of a second assigned opportunity is thus present under every definition of use that we consider, including conversations as short as four minutes.

*[insert Figure 4 around here]*

Taken together, these results provide partial support for H1. The benefits of CAiSEY for voluntary participation are lagged and accrue with a second use: participation is unchanged in the session for which students prepared with CAiSEY, is lower in the next session after a first use, and is higher in the next session and in all future sessions after a second use, under every definition of qualifying use that we consider.

*Self-Reported Learning Experiences*

H2a through H2d predict that using CAiSEY more improves how students experience the course: their comfort with speaking up (H2a), mental focus (H2b), motivation to prepare and participate (H2c), and perceived learning of strategy (H2d). Students rated each aspect on a seven point agreement scale at the midterm and at the end of the term (Supplementary Table S8). The 627 students who completed both surveys form the analysis sample, which spans all ten sections.

Students rated their experience favorably at both points in time, with mean ratings between 5.56 and 5.97 (Figure 5, left; Supplementary Table S10). Comfort with speaking up rose from 5.56 at the midterm to 5.67 at the end of the term, and perceived learning rose from 5.82 to 5.97, whereas focus (5.86 to 5.83) and motivation (5.77 to 5.83) changed little. When looking across students, 28% reported greater comfort and 30% reported greater learning, whereas about half gave the same rating at both points in time (Figure 5, right; Supplementary Table S10).

*[insert Figure 5 around here]*

To relate these changes to CAiSEY, we ask whether the change in a student's ratings from the midterm to the end of the term depends on when the student's two assignments fell: before the midterm survey (sessions 2 through 5) or between the midterm and end-of-term surveys (sessions 7 through 12). We count each student's assignments, or qualifying uses, in each window. Each count ranges from zero to two. Because every student was assigned twice, students with more assignments between the midterm and end of the term had fewer assignments before the midterm (Supplementary Table S9), so the model compares students whose assignments came later in the course with students whose assignments came earlier (Figure 6). We interpret the End-of-term coefficient ($\lambda_E$) as the effect of later use on end-of-term ratings, which assumes that ratings would not otherwise have changed between the surveys.

*[insert Figure 6 around here]*

For comfort speaking up and perceived learning, the effects of CAiSEY emerge by the end of the term. Use between the midterm and the end of the term was associated with greater comfort, whether measured by qualifying use (Supplementary Table S11, Model 2: $\lambda_E$ = 0.131, SE = 0.063, $p < 0.05$) or by assignment (Model 1: $\lambda_E$ = 0.076, SE = 0.035, $p < 0.05$). Perceived learning showed the same pattern (Model 8: $\lambda_E$ = 0.124, SE = 0.055, $p < 0.05$; Model 7: $\lambda_E$ = 0.073, SE = 0.031, $p < 0.05$; Figure 6). Each qualifying use corresponds to roughly one tenth of a midterm standard deviation in either rating (Supplementary Table S10). Coefficients on use before the midterm survey are shown in Figure 6 and reported in Supplementary Table S11.

We find no evidence of corresponding effects for focus or for motivation. Qualifying use between the midterm and the end of the term was not associated with ratings of mental focus (Supplementary Table S11, Model 4: $\lambda_E$ = −0.035, SE = 0.056, $p > 0.1$) or of motivation to prepare and participate (Model 6: $\lambda_E$ = 0.040, SE = 0.066, $p > 0.1$). The four-item index also showed no detectable change (Model 10: $\lambda_E$ = 0.067, SE = 0.044, $p > 0.1$).

As in our analysis of students' voluntary participation, we ask whether these estimates depend on how long a conversation must last to count as qualifying use, with thresholds from five to ten minutes (Figure 7; Supplementary Table S12). End-of-term estimates remain positive and statistically significant at every threshold for comfort ($\lambda_E$ = 0.081 to 0.131; Supplementary Table S12, Panel A, Models 1 to 6; all $p < 0.05$) and perceived learning ($\lambda_E$ = 0.078 to 0.124; Panel D, Models 19 to 24; all $p < 0.05$). Estimated coefficients remain near zero for mental focus (Panel B, Models 7 to 12; all $p > 0.1$) and motivation (Panel C, Models 13 to 18; all $p > 0.1$).

*[insert Figure 7 around here]*

Taken together, these results provide conditional support for H2a and H2d and no support for H2b and H2c. Students whose use came later in the course reported greater comfort speaking up and greater perceived learning at the end of the term. We find no evidence of differences in mental focus or in motivation to prepare and participate. Our interpretation is that the two experiences that are higher with later use, comfort speaking up and perceived learning, also substantiate the argument behind our hypotheses: that less anxiety about speaking up and greater self-efficacy through knowledge mastery are associated with more in-class participation.

## Discussion

We presented results from a field experiment to isolate the treatment effect of using a voice-based AI discussion partner, called CAiSEY, on student learning behaviors and outcomes. Across 759 students enrolled in ten sections of a required course in a large business school, we found that engaging in a real-time voice-based discussion led to greater in-class participation and increases in self-reported comfort with speaking up in class and learning of class material. Random assignment to use CAiSEY for some sessions of the class and not others permitted us to compare students across different levels of exposure to preparation via voice-based AI discussion, allowing for a causal interpretation. Our findings indicate that after using CAiSEY for two sessions, students made 31% more in-class voluntary contributions to class discussion during each session, relative to the sample mean of 0.70 contributions per session.

A recent news feature from *Nature* surfaced ideas for combating the erosion of critical thinking that many educators have attributed to the proliferation of AI services that have allowed students to offload the effortful cognitive processing that is indispensable to learning (Pearson 2026). According to the article, a hallmark of critical thinking is "the ability to consider alternative possibilities or perspectives to our own." Our study of how engaging in a debate-style discussion with voice AI stimulates greater in-class participation sheds light on a potential pathway through which AI can actually enhance the very type of cognitive ability that learners risk losing.

A few limitations constrain the generalizability of our study. First, in terms of study design, we cannot compare the end-of-term learning outcomes of students who used CAiSEY to those who had never used CAiSEY. Because we conducted our study in the field, across multiple sections of a live in-person course, we prioritized fairness in our process of collecting informed consent. This meant that although students were randomized to use CAiSEY for different sessions of class, we were committed to ensuring that every student had the same number of opportunities to use CAiSEY. Second, the interpretation of the positive treatment effect of using CAiSEY on classroom participation only extends to comparing students who used CAiSEY to those that have not yet used CAiSEY. Ideally, we would have liked to compare students who used CAiSEY to engage in a discussion with an AI partner to those who engaged in a pre-class discussion with a human partner. However, enforcement and standardization of having a discussion with a human discussion partner would have created complexity in monitoring and data collection that were beyond the scope of our resources and capacity for attention. Finally, our results are limited to examining a voice-based AI discussion partner that has been purpose-built and tailored to the topics of discussion that students anticipate having in class. We cannot generalize a claim about student use of a generic voice AI chatbot and its effect on their learning outcomes (Bastani et al. 2025).

Our findings make several contributions to the burgeoning conversation about the relationship between student use of AI and their classroom learning outcomes. First, a great deal of the recent literature on AI and learning has focused on assessment outcomes, often examining how students score on standardized tests with the aid of AI tools, such as tailor-built tutors (Kestin et al. 2025, Henkel et al. 2024, generic LLMs (Bastani et al. 2025, Wecks et al. 2024), and other forms of AI augmentation (Wang et al. 2024). As yet, there has been no consensus on whether or how AI might shape student learning outcomes, with some work finding detrimental effects (Bastani et al. 2025, Strömberg, Lei, and Wu 2026) while others finding achievement-enhancing effects (Kestin et al. 2025, De Simone et al. 2025). Our work, instead, focuses on an intermediate learning behavior – in-class participation – which has been documented as a key catalyst for better learning outcomes (Chi and Wylie 2014, Howe et al. 2019). Although we focus on a voice-based AI intervention, ours is the first study to document a causal relationship between pedagogically-directed student AI use and students' class participation behavior.

Furthermore, studies that have assessed the role of AI tools in the classroom, especially the use of LLM-based AI tutors, have almost exclusively looked at texting- or typing-based chat interfaces. We shift the focus to study the role of voice-based AI interactions as a means to help students rehearse and practice their discussion skills. As voice AI becomes increasingly prevalent across various interfaces that touch our daily lives, their use in formal educational contexts is scarce. Our findings represent the first serious large-scale randomized controlled trial to isolate how using a purpose-built voice AI tool can affect student learning outcomes in a classroom setting.

Taken together, our findings are broadly instructive as perspectives from educators have increasingly chronicled how student use of AI has undermined their opportunities for engaging in critical thinking (Gerlich 2025). By offloading their learning activity to generic LLMs, students skip key moments characterized by learning frictions that are central to developing a capacity for critical thinking. When students forgo these opportunities, they also lose motivation to engage in further learning opportunities, such as sharing their ideas and perspectives in classroom discussion. Our findings point to voice AI platforms as a way for instructors to engage their students in new modalities of learning that can not only trigger greater critical thinking, but also motivate students to participate in other vital learning-based interactions, such as classroom discussion.

**Methods**

*Ethics.* This study was approved by the Institutional Review Board (IRB) of Columbia University (Protocol AAAV9941) and classified as exempt. All methods were carried out in accordance with relevant guidelines and regulations. Informed consent was obtained from all subjects; students were provided with an information sheet detailing the study and consented to the analysis of their data, retaining the freedom to opt out at any time.

*Setting and experimental design.* The experiment was conducted in a required first-year MBA strategy course at a US business school during Fall 2025. The course used the case method: students read and analyzed a business case before class and discussed competing interpretations in an instructor-led meeting. Ten sections, each containing 72 to 77 students, followed the same case calendar and were taught by four instructors. Students were placed in sections administratively by the school, and 759 students were randomized in total (Supplementary Table S2, Panel B).

The course comprised twelve meetings, of which ten used CAiSEY: sessions 2 to 5 and 7 to 12. Session 1 was the introductory meeting, and session 6 involved group presentations that did not lend themselves to the same type of structured case discussions in the other sessions. Before the course, each student was randomly assigned two of the ten eligible cases. Randomization was stratified by section to distribute assignments across the case calendar within each section. All students therefore received two assigned opportunities, and the timing of those opportunities varied across students (Figure 1; Supplementary Table S1).

Students were informed by email of their assigned cases at the start of the course and received email reminders about using CAiSEY before the class sessions to which they were assigned. Instructors and Teaching Assistants were blind to student assignment and usage. Teaching assistants recorded participation in class by noting each instance of a student speaking up. They also noted whether a student was 'cold called' – i.e., called upon by the instructor to speak when the student's hand was not raised – which allowed us to measure a student's voluntary contributions to class discussion.

Midterm and end-of-term surveys were administered online after sessions 6 and 12, respectively. Participation records, platform logs and survey responses were linked using student identifiers. A pre-course poll measured comfort participating in class before any CAiSEY use, and student resumes provided students' years of work experience. We use both measures to assess balance across assignment timing (Supplementary Table S14).

*The CAiSEY intervention.* When a student was assigned to use CAiSEY, the interaction proceeded as follows. Each case came with an initial discussion question. For example, "Does a publicly owned firm have an obligation to consider sustainability in striving toward its financial goals?" The student chose a position on the discussion question and CAiSEY automatically adopted an opposing position. The student opened the conversation by defending their chosen stance, then CAiSEY responded by challenging the student's arguments and raising alternative perspectives in a spoken exchange. This proceeded in alternating turns until the student chose to end the conversation. The student then submitted a written reflection and received an automatically generated summary and feedback. Instructions directed students to read the case before completing the exercise and to finish before class. The platform recorded the student identifier, case, submission timestamp, conversation duration, positions taken, transcript, reflection and generated summary.

*Class participation outcomes.* At each class meeting, a teaching assistant recorded attendance, the number of times each student was recognized to speak and the number of those contributions initiated by an instructor cold call. The primary behavioral outcome was the voluntary contribution count, defined as recorded contributions minus cold calls, which counts the contributions that students chose to make in the discussion.

The participation panel covered all twelve sessions, including the midterm meeting (session 6) as a session without a CAiSEY assignment. Session 1 enters only as the source of prior session controls for session 2. Of the 759 randomized students, 3 lacked roster or participation records, and two sections containing 152 students did not record cold calls separately, so voluntary contributions could not be distinguished from instructor-initiated participation. We also excluded student-sessions in which attendance was recorded as absent (716 student-sessions) and those in which the cold-call count exceeded the total contribution count (11 student-sessions) (Supplementary Table S2, Panel B).

In one retained section, attendance was recorded as absent for all students in sessions 2 to 11. We reconstructed attendance as present when a contribution was recorded and retained only contributing observations for those meetings. Because inclusion in this section depends on the outcome, section-by-session fixed effects do not fully address it, so we also estimate the participation models without this section (Supplementary Table S15). The resulting panel contains 6,003 student-sessions from 604 students in eight sections. Exposure-specific models were estimated among the 504 students whose two assigned sessions were observed, regardless of whether they completed the assigned conversations (5,291 student-sessions). Requiring the previous observed session's outcome reduced this sample to 4,787 observations, and the pooled Same Session assignment model used 5,395 observations from 596 students (Supplementary Table S2, Panel B). Excluding the section with reconstructed attendance leaves the pattern of results unchanged: the second assigned opportunity remains associated with more participation in the next session and in all future sessions (Supplementary Table S15, Models 4 to 7: $\beta_2$ = 0.101 to 0.213; all $p < 0.05$), and Same Session estimates remain close to zero (Models 2 and 3; both $p > 0.1$).

*Self-reported learning experience outcomes.* Both surveys measured four aspects of the learning experience: comfort speaking up and sharing a perspective; mental focus on class discussion; motivation to prepare and participate; and gaining useful insights that deepened understanding of strategy. Responses used a seven-point agreement scale, from 1 (strongly disagree) to 7 (strongly agree), with 4 as the midpoint. The items correspond to H2a through H2d, respectively. We analyzed each item separately and also constructed a summary index as their unweighted mean. These measures capture students' perceptions of their learning experience. The verbatim items and response anchors are in Supplementary Table S8.

The midterm survey was returned by 690 students and the end-of-term survey by 678; 627 returned both. Returning both surveys did not differ by assignment timing (Supplementary Table S14, Panel B: joint test, $p > 0.1$). The primary survey panel retained respondents to both surveys across all ten sections, because survey measurement did not depend on the availability of cold-call records. A secondary panel retained the 503 respondents who also belonged to the participation cohort. The number of available responses differed slightly across items (Supplementary Tables S10 and S11).

*Assignment and qualifying use.* An assignment indicator identified whether the case in a given session was one of the student's two randomly assigned cases. The first and second assigned opportunity denote the earlier and later of a student's two assigned sessions. Recorded use indicated at least one submission for the assigned case. Qualifying use required a conversation lasting at least ten minutes, and this threshold was used for the main participation and survey analyses. If a student submitted more than one conversation for the same case, we kept the longest.

Among recorded conversations in the participation estimation sample, duration averaged 11.27 minutes (SD = 7.21; n = 851; Supplementary Table S3). Assignment strongly predicted qualifying use at the first and second assigned opportunity, with first-stage coefficients of 0.492 (SE = 0.025) and 0.527 (SE = 0.022), respectively (Supplementary Table S5, Panel A, Models 13 and 14).
For the survey analysis, assigned and qualifying sessions were counted separately before the midterm (sessions 2 to 5) and afterwards (sessions 7 to 12). Each assigned count ranged from zero to two, and the two counts summed to two for every student (Supplementary Table S9).

*Class participation models.* Linear models were estimated on the student-session panel. For student $i$ in section $c$ at session $s$, the exposure-specific specification was:

$$Y_{is} = \alpha_i + \gamma_{cs} + \beta_1 X_{1,is} + \beta_2 X_{2,is} + \rho Y_{i,s-1} + \varepsilon_{is} \quad (1)$$

where $Y_{is}$ is the number of voluntary contributions by student $i$ in session $s$; $\alpha_i$ are student fixed effects; $\gamma_{cs}$ are section-by-session fixed effects; $X_{1,is}$ and $X_{2,is}$ indicate exposure to the first and second assigned opportunity (assignment in the ITT models and qualifying use in the IV models) in the relevant timing window; and $Y_{i,s-1}$ is the student's voluntary contribution count in the previous observed session.

Participation varied considerably across class meetings (Supplementary Figure S1). Student fixed effects account for stable individual differences in participation. Section-by-session fixed effects allow the participation baseline to vary for every section meeting, absorbing shared differences in the case, instructor and discussion format. The lagged outcome is the voluntary contribution count in the student's previous observed session, which is not always the previous scheduled meeting. Coefficients are expressed in voluntary contributions per session.

The two exposure variables were defined in three ways. Same Session indicators identified the assigned discussion itself. Next Session indicators identified the first observed meeting after each exposure. All Future Sessions indicators equaled one in every observed meeting after the corresponding exposure, so that both indicators are active after the second exposure. In this model the coefficient on the second exposure ($\beta_2$) is the increment over the coefficient on the first ($\beta_1$), and their sum is the combined post-exposure effect (Supplementary Table S4, Models 6 and 7). The three timing specifications were estimated separately, and the pooled Same Session ITT model replaced the two exposure indicators with a single assignment indicator with coefficient β (Supplementary Table S4, Model 1).

ITT models used the assignment indicators. Two-stage least squares models replaced these with qualifying use indicators, instrumented by the corresponding randomized assignment indicators, and retained the same fixed effects and lagged outcome control. Interpreting the IV coefficients as local average treatment effects requires relevant instruments, monotonicity, an exclusion restriction and a correct specification of treatment history. The exclusion restriction requires that assignment affects participation only through qualifying use. Conversations shorter than ten minutes, reminders and changes in other preparation are potential departures from it, which is one reason we vary the duration threshold (Figure 4). The first stage is strong at every threshold (Supplementary Table S5, Panel A: coefficients 0.49 to 0.92, all $p < 0.001$).

For the illustration in Figure 3, predictions from a model without the lagged outcome control compared assignment in sessions 3 and 5 with a hypothetical assignment in session 3 alone. Predictions used mean student and section-by-session fixed effects. Differences were summed over sessions 5 through 12 to give the cumulative contrast (Δ) in Supplementary Table S6, and session-level predictions are in Supplementary Table S7. Confidence intervals were calculated from the covariance of the estimated coefficients. Because every student received two assignments, the one opportunity path is a model-based counterfactual.

*Self-reported learning experience models.* The survey panel contained a midterm and an end-of-term row for each respondent. The two dose variables (Midterm and End-of-term) count assigned or qualifying sessions before the midterm survey (sessions 2 through 5) and between the midterm and end-of-term surveys (sessions 7 through 12), and each count entered only its corresponding survey observation. Outcomes and regressors were expressed as deviations from the student's two-wave mean, which for complete pairs is algebraically equivalent to differencing the two observations and removes stable differences in response levels. The model was:

$$Y_{it} = \alpha_i + \lambda_M D_{iM}\, I(t = M) + \lambda_E D_{iE}\, I(t = E) + \varepsilon_{it} \quad (2)$$

where $Y_{it}$ is student $i$'s rating at wave $t$ (midterm $M$ or end of term $E$); $\alpha_i$ are student fixed effects; $D_{iM}$ and $D_{iE}$ are the numbers of assigned (ITT) or qualifying (IV) sessions of student $i$ before the midterm survey and between the midterm and end-of-term surveys, respectively; and $I(t = M)$ and $I(t = E)$ indicate the survey wave.

In the IV models, the wave-specific qualifying counts were instrumented by their assigned-count counterparts. The Midterm coefficient ($\lambda_M$) pertains to the midterm rating and the End-of-term coefficient ($\lambda_E$) to the end-of-term rating, and the model has no separate carryover term. Because the two assigned counts sum to two for every student, interpreting $\lambda_E$ as the effect of later use requires that ratings would not otherwise have changed between the surveys and that earlier use does not affect the end-of-term rating. We report the coefficients for both windows in Supplementary Table S11.

*Statistical inference and sensitivity analyses.* Standard errors were clustered by student in the participation and survey analyses. Coefficients are reported in the original outcome units: voluntary contributions per student-session and scale points for the surveys. Tests are two-sided and are not adjusted for multiple comparisons. We report significance at the 10%, 5%, 1% and 0.1% levels, and the 95% confidence intervals in the figures use a normal approximation.

Sensitivity analyses varied the threshold that defines qualifying use (four to ten minutes for participation and five to ten minutes for the surveys), restricted survey respondents to the participation cohort, and excluded the participation section with reconstructed attendance. Threshold analyses redefine both the treatment and the set of students who qualify at each cutoff while holding random assignment fixed (Supplementary Tables S5 and S12). Among the 503 survey respondents who also appear in the participation sample, the End-of-term estimates are similar in size for comfort (Supplementary Table S13, Model 2: $\lambda_E$ = 0.135, SE = 0.077, p < 0.1) and perceived learning (Model 8: $\lambda_E$ = 0.120, SE = 0.066, p < 0.1), though less precisely estimated.

*Protocol registration.* The protocol was registered on October 2, 2025 and is available at https://osf.io/ypt74.

**Data and code availability.** Anonymized data and code to replicate all findings will be made available in a persistent repository upon acceptance for publication.

**Acknowledgments.** This project was generously supported by the AI in Business Initiative at Columbia Business School. We thank Jill Cohen and Johnny Lee for helping to develop CAiSEY and the MBA and EMBA students at Columbia Business School who opted to use CAiSEY and provided valuable feedback.

**Author Contributions.** D.J.W. designed research, performed research, analyzed data, and wrote the paper; N.M.J. analyzed data and wrote the paper; V.B., J.G., D.K., S.K., B.K., and N.W. facilitated data collection.

**Materials & Correspondence.** Correspondence and material requests to Dan J. Wang at djw2104@gsb.columbia.edu.

**Competing Interests.** The authors declare no competing interest.

**Funding.** This research received no external funding.

## Figures

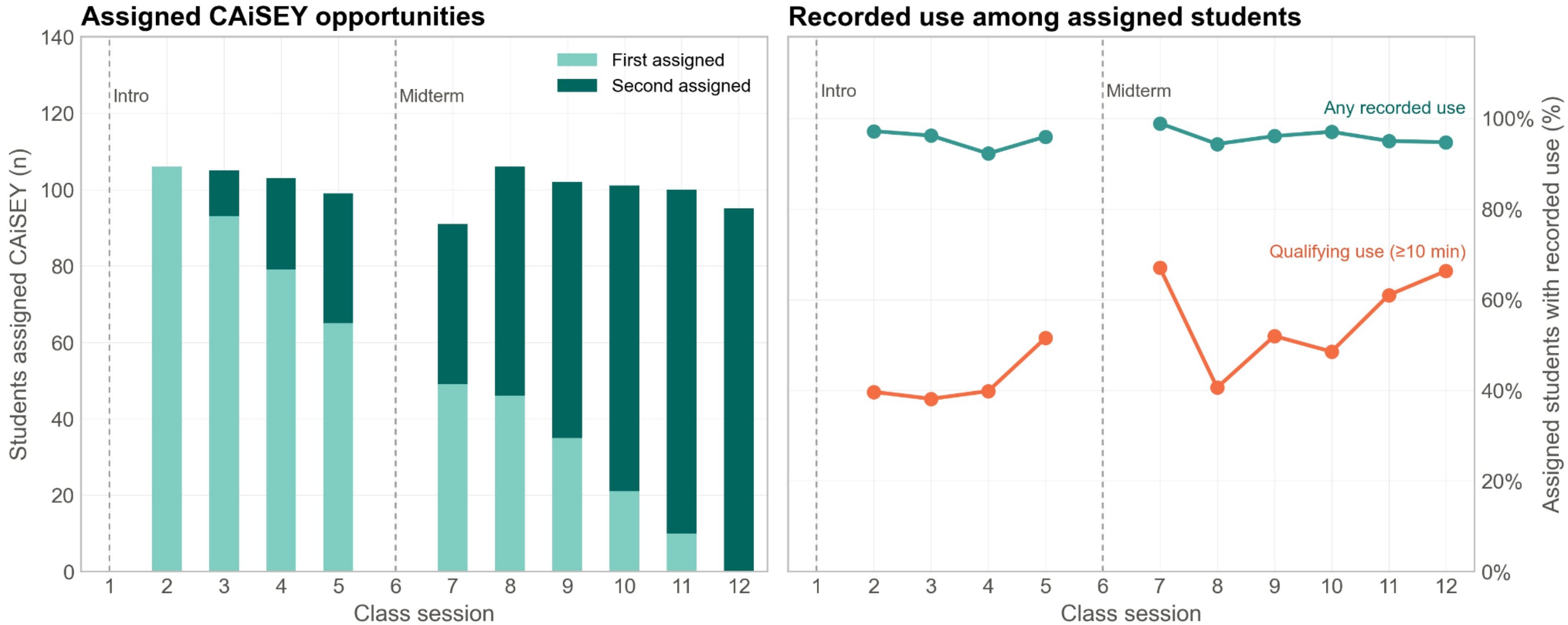

**Figure 1. Assignment to CAiSEY and recorded use across class sessions.** Left: Number of students assigned their first or second CAiSEY opportunity in each class session. Right: Percentage of assigned students with any recorded use and with qualifying use, defined as an interaction lasting at least ten minutes. The denominator is the number of students assigned to CAiSEY in that session. Each student received two assigned opportunities, and the sample is the 504 students whose two assigned sessions were observed in the participation panel. Sessions 1 and 6 had no assignments, so use percentages are not plotted for those sessions. Counts and denominators are reported in Supplementary Table S1.

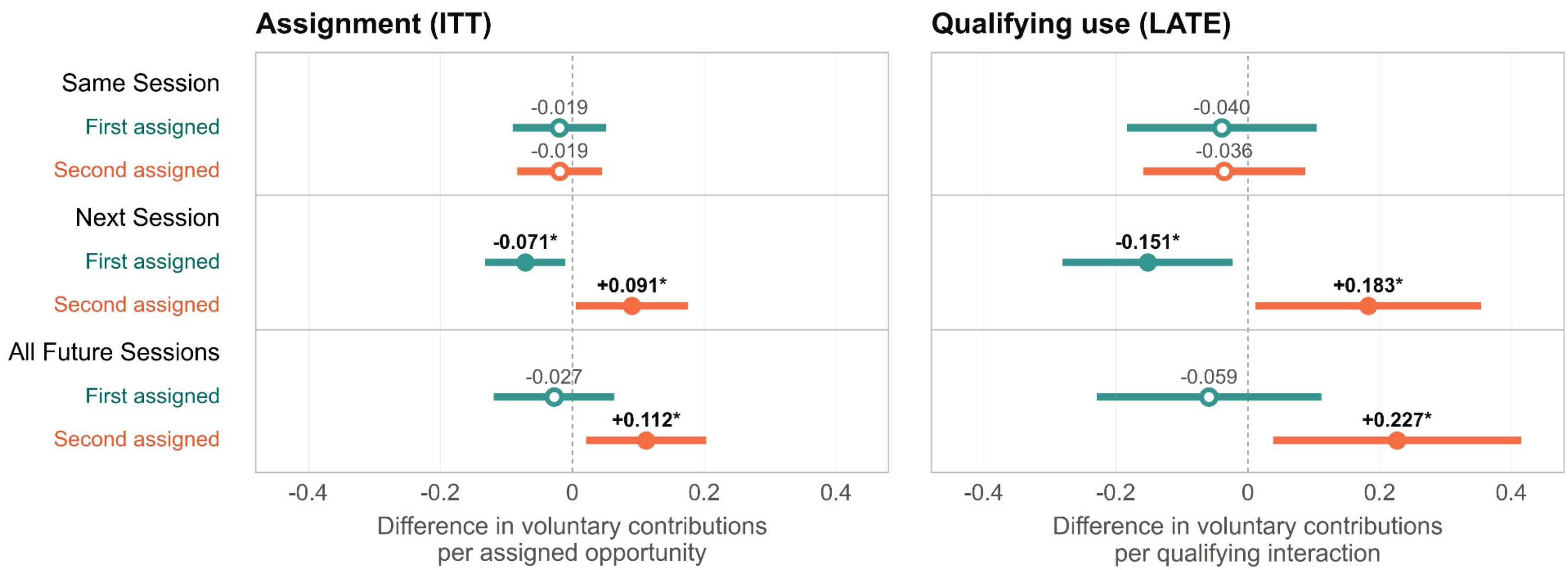

**Figure 2. Class participation estimates by assigned opportunity and outcome timing.** Left: Intention-to-treat (ITT) estimates of assignment. Right: Instrumental-variable estimates of qualifying use induced by assignment, interpreted as local average treatment effects (LATEs) under the assumptions in Methods. Qualifying use requires an interaction of at least ten minutes. Same Session refers to the assigned class meeting, Next Session to the first observed meeting afterwards, and All Future Sessions to all subsequent observed meetings. First and Second identify the earlier and later of a student's two assigned opportunities. In All Future Sessions, the first assigned opportunity term remains active after the second assigned opportunity, so the second coefficient is the additional effect of the second assigned opportunity and the combined effect is the sum of the two. Circles show coefficients, and horizontal lines show 95% confidence intervals. Filled circles indicate $p < 0.05$; open circles indicate $p \geq 0.05$. Models include student fixed effects, section-by-session fixed effects and prior observed participation, with standard errors clustered by student, and use 4,787 student-session observations from 504 students. Full results are in Supplementary Table S4 (Models 2 to 7). Significance levels: +p < 0.1, *p < 0.05, **p < 0.01, ***p < 0.001.

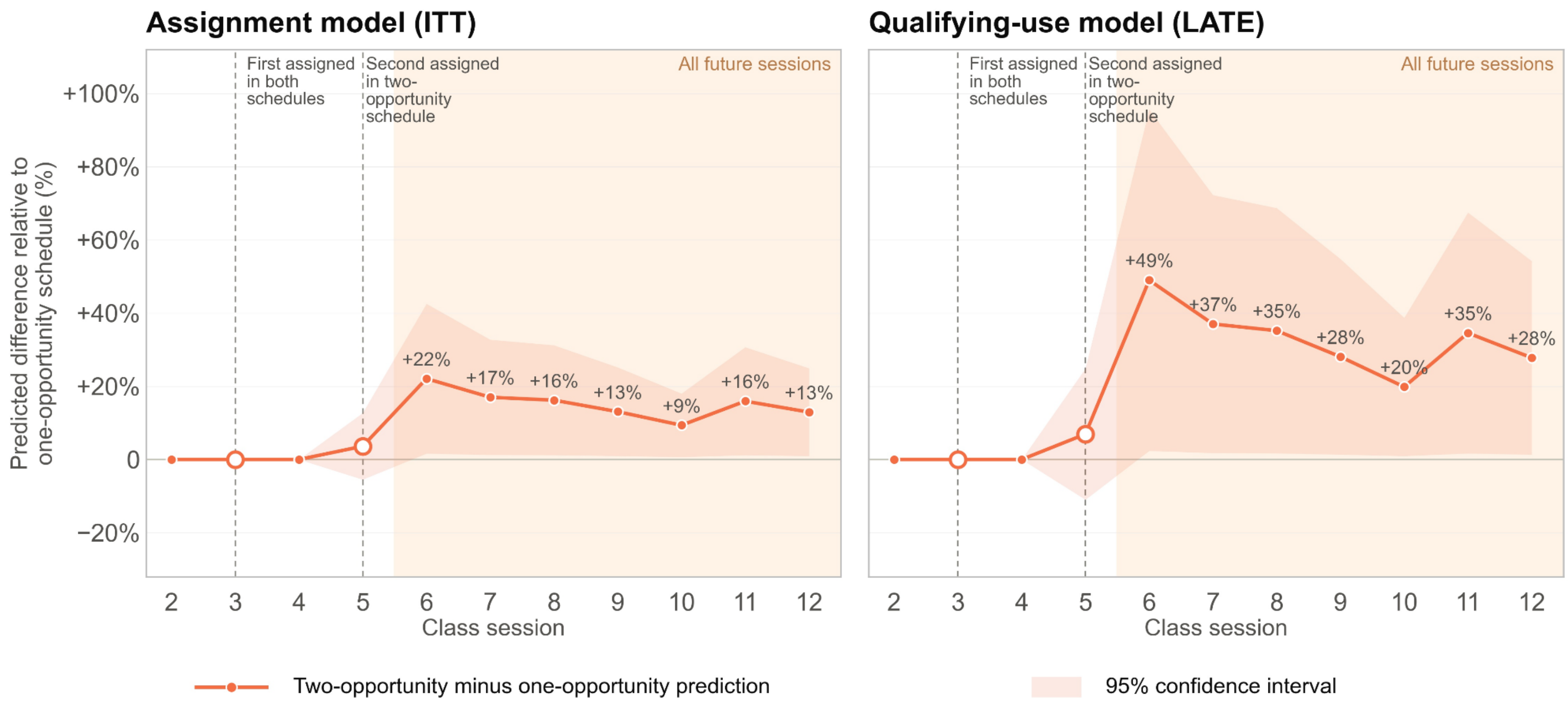

**Figure 3. Model-predicted participation with and without a second assigned opportunity.** Predictions compare assignment in sessions 3 and 5 with a hypothetical schedule containing assignment in session 3 alone. Left, assignment-based predictions (ITT). Right, qualifying use predictions from the instrumental-variable model (LATE). The vertical axis is the predicted difference divided by the one-opportunity prediction in the same class session, expressed as a percentage. Both schedules include the opportunity in session 3, and only the two-opportunity schedule includes session 5. Shading denotes 95% confidence intervals. Predictions use mean student and section-by-session fixed effects from a specification without the prior-participation control used in Figure 2. Because every student received two assignments, the one-opportunity path is a model counterfactual. Percentage differences depend on the reference level of participation in each session. Numerical contrasts are in Supplementary Table S6, and the prediction values underlying the figure are in Supplementary Table S7.

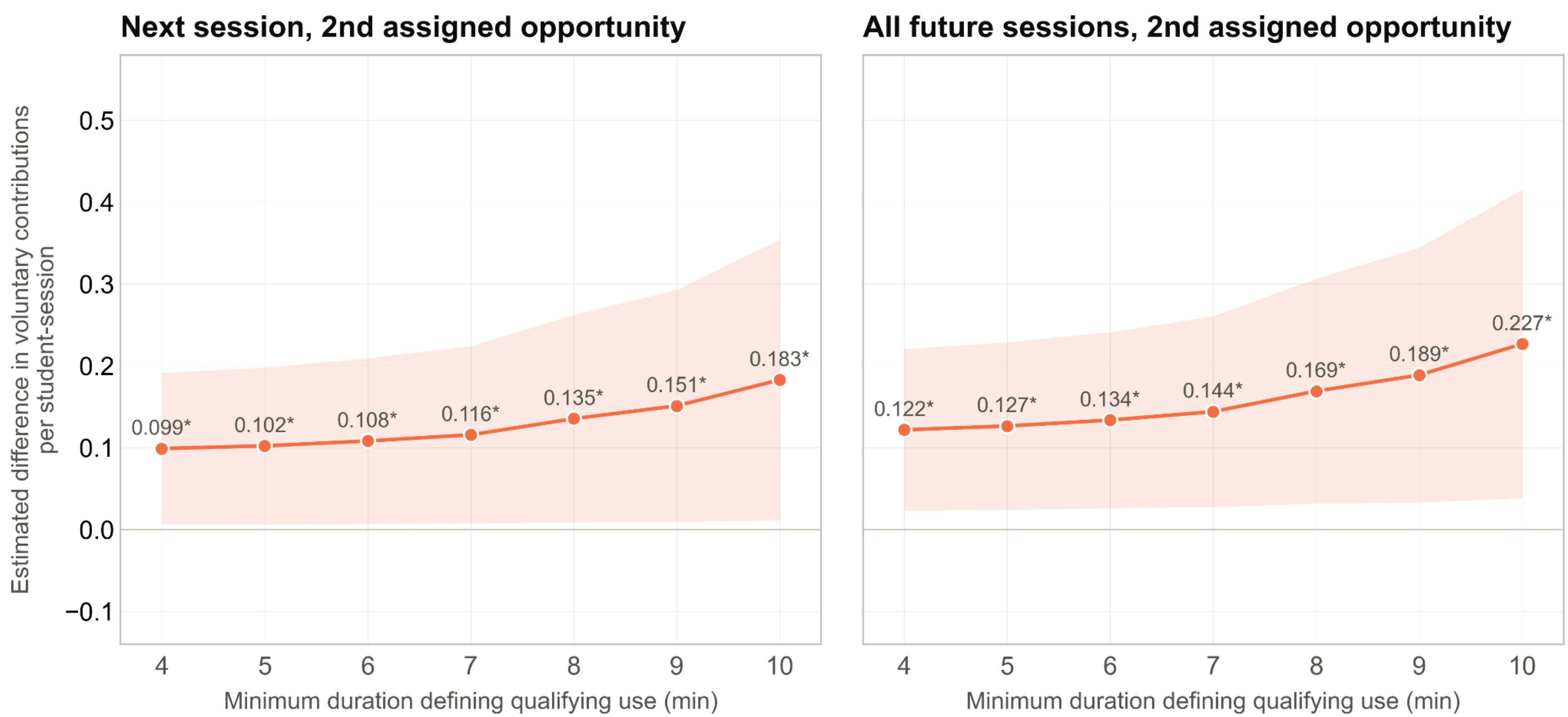

**Figure 4. Participation estimates across qualifying use duration thresholds.** Instrumental-variable estimates for the second assigned opportunity are shown for Next Session (left) and All Future Sessions (right). In All Future Sessions, the coefficient is the additional second assigned opportunity term. Each threshold redefines qualifying use as an interaction lasting at least the indicated number of minutes. Circles show coefficients and shaded intervals show 95% confidence intervals; point annotations give the coefficient, and asterisks denote $p < 0.05$. Models include student and section-by-session fixed effects and prior observed participation, with standard errors clustered by student, and use 4,787 student-session observations from 504 students. Full estimates are in Supplementary Table S5, Panels B and C, and first-stage estimates are in Panel A.

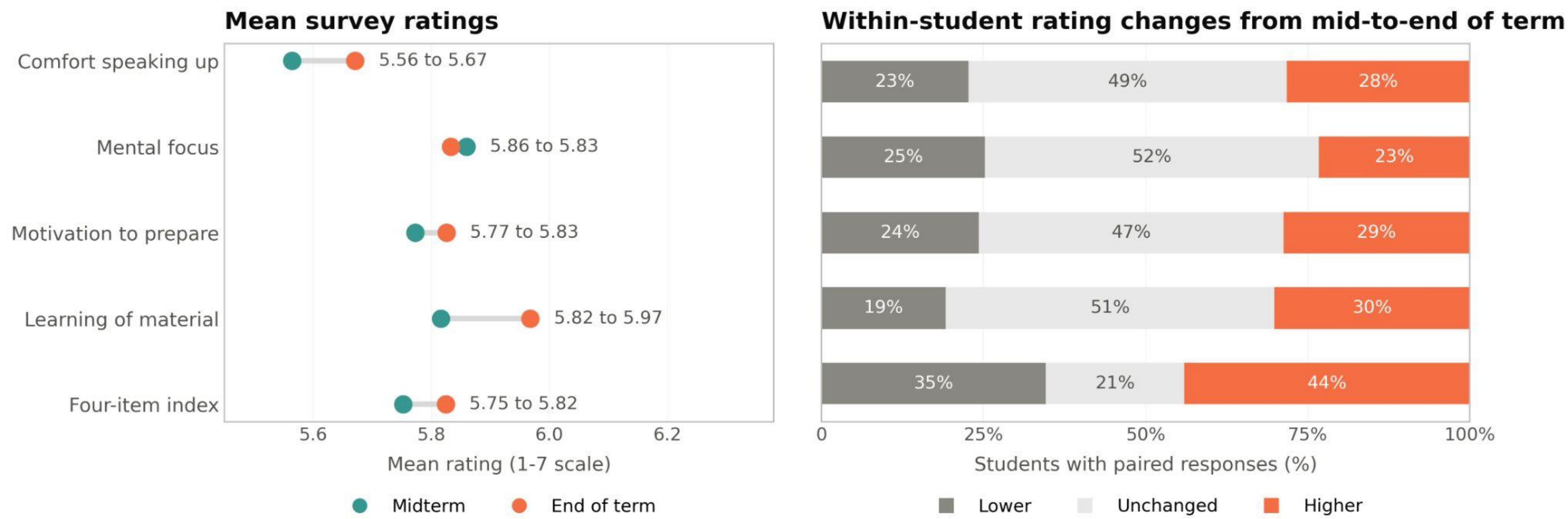

**Figure 5. Self-reported learning experiences at midterm and end of term.** Left: Mean ratings at midterm and end of term. Right: Percentages of students whose end-of-term rating was lower than, equal to, or higher than their midterm rating. Responses range from 1 (strongly disagree) to 7 (strongly agree). The four-item index is the unweighted mean of the four ratings. The sample comprises 627 students who returned both questionnaires, and the number of available responses and complete pairs differs slightly by item. Item-specific sample sizes, standard deviations and change percentages are reported in Supplementary Table S10.

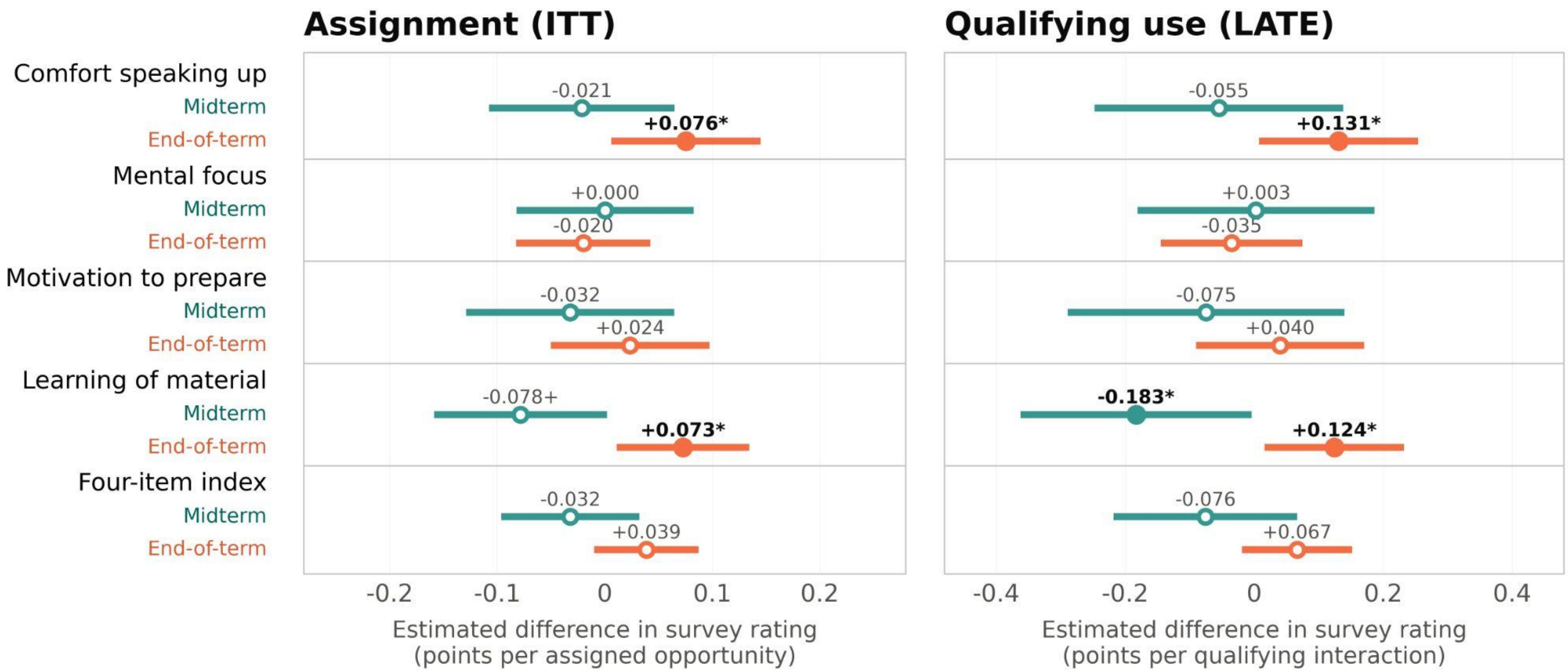

**Figure 6. CAiSEY assignment and qualifying use estimates for self-reported learning experiences.** Left: Intention-to-treat (ITT) coefficients per assigned CAiSEY opportunity. Right: Instrumental-variable coefficients per qualifying interaction induced by assignment, reported as local average treatment effects (LATEs) under the assumptions in Methods. Qualifying use requires an interaction of at least ten minutes. Assignments or uses in sessions 2 to 5 enter the midterm observation (Midterm rows), and those in sessions 7 to 12 enter the end-of-term observation (End-of-term rows). These labels refer to course timing, not to a student's first and second assigned opportunities. Circles show coefficients in points on the seven-point response scale, and horizontal lines show 95% confidence intervals. Filled circles indicate $p < 0.05$; open circles indicate $p \geq 0.05$. Variables are demeaned within student, and standard errors are clustered by student. The sample contains 627 respondents, with 1,254 student-wave observations for comfort and the index, 1,253 for learning and motivation, and 1,252 for focus. Estimates of the separate Midterm and End-of-term coefficients rest on the temporal and carryover assumptions described in Methods. Full results are in Supplementary Table S11. Significance levels: +p < 0.1, *p < 0.05, **p < 0.01, ***p < 0.001.

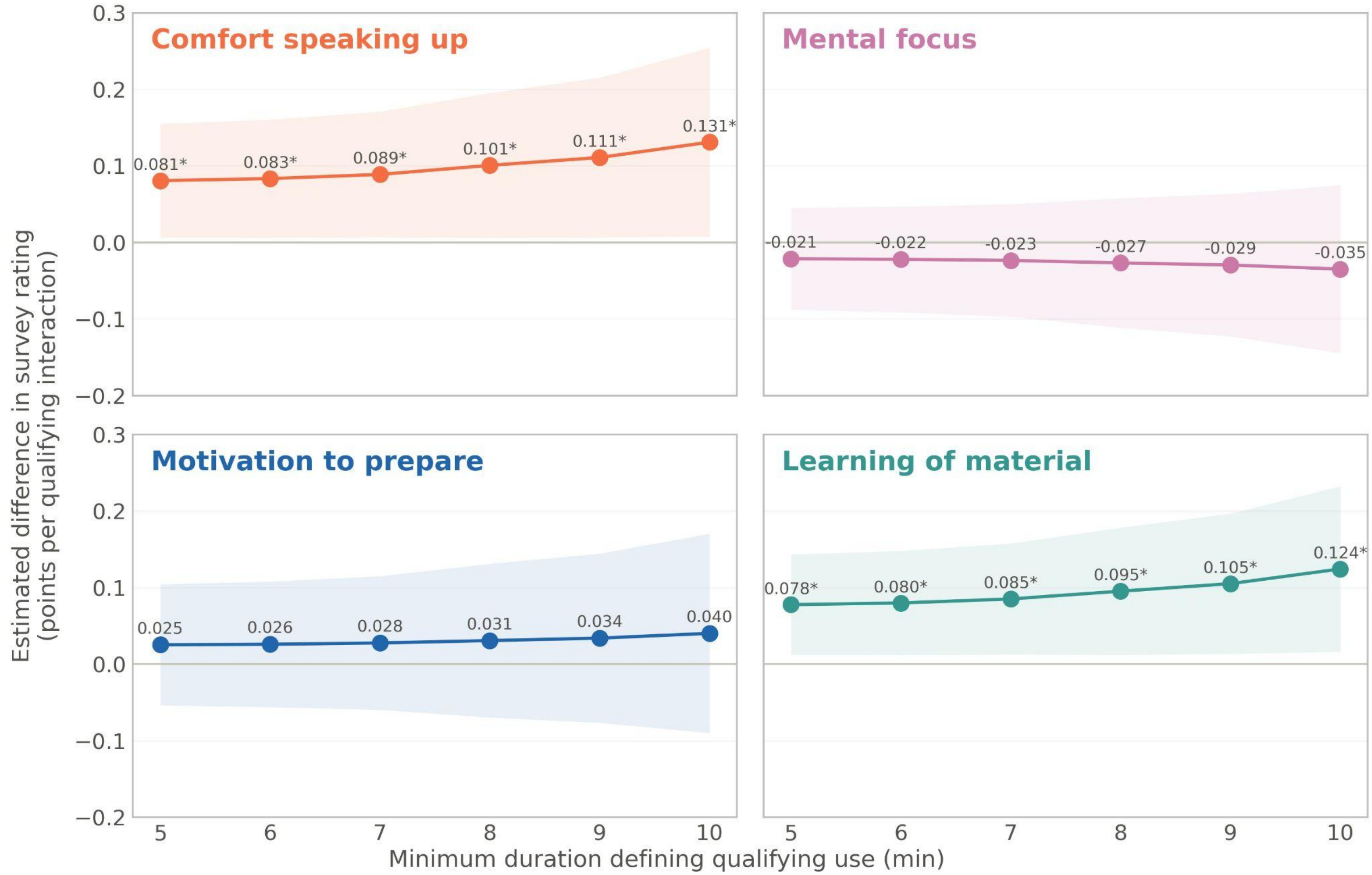

**Figure 7. End-of-term survey estimates across qualifying use duration thresholds.** End-of-term instrumental-variable coefficients are shown for comfort speaking up, mental focus, motivation to prepare and participate, and perceived learning of material. Each threshold defines qualifying use as an interaction lasting at least the indicated number of minutes. Circles show estimated differences in survey ratings per qualifying interaction, shaded intervals show 95% confidence intervals, and asterisks denote $p < 0.05$. Variables are demeaned within student, and standard errors are clustered by student. The sample contains 627 respondents, with 1,254 student-wave observations for comfort, 1,252 for focus and 1,253 for learning and motivation. Models keep the Midterm and End-of-term specification of Figure 6. Full estimates, including the four-item index, are in Supplementary Table S12.

## Supplementary Information

# How assigned AI use before class shapes active student engagement in class

Dan J. Wang, Neelam Modi Jain, Vanessa Burbano, Jorge Guzman, Daniel Keum, Soomi Kim, Bruce Kogut, Nataliya Wright

**Contents**

**Table S1.** CAiSEY assignment and recorded use by class session

| Class session | First assigned opportunity | Second assigned opportunity | All assigned opportunities | Any recorded use | Qualifying use | Any use / assigned use | Qualifying use / assigned use | Observed students |
|---|---|---|---|---|---|---|---|---|
| # | N | N | N | N | N | % | % | N |
| 1 | 0 | 0 | 0 | N/A | N/A | N/A | N/A | 480 |
| 2 | 106 | 0 | 106 | 103 | 42 | 97.2 | 39.6 | 491 |
| 3 | 93 | 12 | 105 | 101 | 40 | 96.2 | 38.1 | 492 |
| 4 | 79 | 24 | 103 | 95 | 41 | 92.2 | 39.8 | 499 |
| 5 | 65 | 34 | 99 | 95 | 51 | 96.0 | 51.5 | 494 |
| 6 | 0 | 0 | 0 | N/A | N/A | N/A | N/A | 484 |
| 7 | 49 | 42 | 91 | 90 | 61 | 98.9 | 67.0 | 480 |
| 8 | 46 | 60 | 106 | 100 | 43 | 94.3 | 40.6 | 474 |
| 9 | 35 | 67 | 102 | 98 | 53 | 96.1 | 52.0 | 487 |
| 10 | 21 | 80 | 101 | 98 | 49 | 97.0 | 48.5 | 457 |
| 11 | 10 | 90 | 100 | 95 | 61 | 95.0 | 61.0 | 463 |
| 12 | 0 | 95 | 95 | 90 | 63 | 94.7 | 66.3 | 470 |

*Notes:* Qualifying use is an interaction of at least ten minutes. Use rates divide by the number of students assigned in that session. Sessions 1 and 6 had no assignments, so use rates are not defined. Observed students is the number of participation panel rows in that session and is not the denominator of the use rates. First and second assigned opportunities each sum to 504 students, and all assigned opportunities sum to 1,008. Session 1 observations serve only to construct prior session controls, and the 5,291 two-exposure student-sessions cover sessions 2 through 12. The assignment allocation of the survey sample is in Supplementary Table S9.

**Table S2.** Class participation sample construction and descriptive statistics

**Panel A. Estimation sample descriptive statistics**

| Sample | Measure | N | Mean | SD | Min | Max |
|---|---|---|---|---|---|---|
| Pooled ITT | Voluntary contributions | 5,395 | 0.757 | 0.847 | 0 | 6 |
| | All recorded contributions | 5,395 | 0.781 | 0.850 | 0 | 6 |
| | Instructor-initiated cold calls | 5,395 | 0.024 | 0.154 | 0 | 1 |
| Opportunity-specific | Voluntary contributions | 4,787 | 0.727 | 0.831 | 0 | 6 |
| | All recorded contributions | 4,787 | 0.750 | 0.835 | 0 | 6 |
| | Instructor-initiated cold calls | 4,787 | 0.023 | 0.149 | 0 | 1 |

**Panel B. Sample Construction**

| Stage or Exclusion | Count |
|---|---|
| Randomized enrollment | 759 students |
| Students without roster or participation records | 3 students |
| Excluded through attendance screening | 716 student-sessions |
| Two sections without separate cold-call records | 152 students |
| Records with more cold calls than contributions | 11 student-sessions |
| Retained analytic panel | 604 students; 6,003 student-sessions |
| Both assigned sessions recorded | 504 students; 5,291 student-sessions |
| Opportunity-specific model sample requiring prior participation | 4,787 student-sessions |
| Pooled Same Session ITT sample | 596 students; 5,395 student-sessions |

*Notes:* Descriptives in Panel A are counts per student-session. Voluntary contributions equal all recorded contributions minus instructor-initiated cold calls. At least one voluntary contribution was recorded in 52.9% of the 4,787 observations of the opportunity-specific sample. The retained analytic panel covers eight sections. Attendance screening dropped student-sessions that remained coded absent after imputation. In one section, attendance was recorded as absent for every student in sessions 2–11; those section-sessions were retained only if at least one contribution was recorded, leaving 402 analytic rows (163 in the opportunity-specific sample) whose raw attendance field is still 0 (Methods). All other absences were excluded. SD is the sample standard deviation. Regression results are in Supplementary Table S4.

**Table S3.** Duration of CAiSEY conversations by analysis sample

| **Sample** | **Any recorded use** | | | | | **Qualifying use** | | | | | |
|---|---|---|---|---|---|---|---|---|---|---|---|
| | **Unique conversations** | **Unique students** | **Duration** (in minutes) | | | **Share of any recorded use** | **Unique conversations** | **Unique students** | **Duration** (in minutes) | | |
| | **N** | **N** | **Mean** | **SD** | **Median** | **%** | **N** | **N** | **Mean** | **SD** | **Median** |
| Participation panel | 851 | 493 | 11.27 | 7.21 | 10.32 | 54.3 | 462 | 307 | 14.43 | 8.43 | 13.03 |
| Survey full panel | 1,194 | 622 | 11.81 | 20.96 | 10.33 | 54.3 | 648 | 414 | 15.44 | 27.90 | 13.03 |
| Survey participation panel | 955 | 500 | 12.02 | 23.33 | 10.28 | 54.0 | 516 | 327 | 15.84 | 31.20 | 13.04 |

*Notes:* Qualifying use is an interaction of at least ten minutes. Each row describes recorded conversations on assigned cases. Participation panel: recorded submissions with non-missing duration in the 4,787 student-session rows used for the opportunity-specific participation models (Supplementary Table S2). Survey full panel: the 627 students who returned both surveys, across ten sections. Survey participation panel: the 503 of those students who are also in the participation analytic sample (Supplementary Table S13). A few recorded durations are extreme, including one of about 699 minutes, so the median is the more informative summary of a typical conversation.

**Table S4.** Effects of CAiSEY on voluntary class participation by assigned opportunity and outcome timing

| **Dependent Variable** | **Voluntary contributions** | | | | | | |
|---|---|---|---|---|---|---|---|
| | **Same session** | | | **Next session** | | **All future sessions** | |
| | Pooled ITT | ITT | LATE | ITT | LATE | ITT | LATE |
| | (1) | (2) | (3) | (4) | (5) | (6) | (7) |
| Assignment (β) | -0.011 | | | | | | |
| | (0.025) | | | | | | |
| First assigned opportunity ($\beta_1$) | | -0.019 | -0.040 | -0.071* | -0.151* | -0.027 | -0.059 |
| | | (0.036) | (0.073) | (0.031) | (0.066) | (0.047) | (0.087) |
| Second assigned opportunity ($\beta_2$) | | -0.019 | -0.036 | 0.091* | 0.183* | 0.112* | 0.227* |
| | | (0.033) | (0.062) | (0.043) | (0.087) | (0.046) | (0.096) |
| First plus second ($\beta_1 + \beta_2$) | | | | | | 0.085 | 0.168 |
| | | | | | | (0.071) | (0.140) |
| Prior voluntary contributions ($\rho$) | -0.100*** | -0.102*** | -0.102*** | -0.102*** | -0.102*** | -0.103*** | -0.103*** |
| | (0.015) | (0.016) | (0.016) | (0.016) | (0.016) | (0.016) | (0.016) |
| Student FE | Yes | Yes | Yes | Yes | Yes | Yes | Yes |
| Section-by-session FE | Yes | Yes | Yes | Yes | Yes | Yes | Yes |
| Clustered SE | Student | Student | Student | Student | Student | Student | Student |
| Observations | 5,395 | 4,787 | 4,787 | 4,787 | 4,787 | 4,787 | 4,787 |
| Students | 596 | 504 | 504 | 504 | 504 | 504 | 504 |
| $R^2$ | 0.422 | 0.398 | 0.398 | 0.400 | 0.400 | 0.399 | 0.399 |

*Notes:* ITT columns report the effect of assignment. LATE columns report the effect of qualifying use, defined as a conversation of at least ten minutes, instrumented by assignment. $\beta_1$ and $\beta_2$ are the coefficients on the first and second assigned opportunity in Equation 1, and β is the coefficient on the single assignment indicator in Model 1. In Models 2 through 7, the first and second assigned opportunity coefficients come from the same regression. In Models 6 and 7 the second assigned opportunity coefficient is the increment beyond the first, and the first plus second row reports their sum ($\beta_1 + \beta_2$) with its standard error. Standard errors are in parentheses. Every model includes student fixed effects, section-by-session fixed effects and prior voluntary contributions, with standard errors clustered by student. Models 2 through 7 use the 504-student opportunity-specific sample, and Model 1 uses the pooled Same Session sample (Supplementary Table S2, Panel B). Significance levels: +p < 0.1, *p < 0.05, **p < 0.01, ***p < 0.001.

**Table S5.** Effects of qualifying use on voluntary class participation across duration thresholds

**Panel A. First stage: effect of assignment on qualifying use**

| **Dependent variable** | **Qualifying use** | | | | | | | | | | | | | |
|---|---|---|---|---|---|---|---|---|---|---|---|---|---|---|
| | **4 min** | | **5 min** | | **6 min** | | **7 min** | | **8 min** | | **9 min** | | **10 min** | |
| | First | Second | First | Second | First | Second | First | Second | First | Second | First | Second | First | Second |
| | (1) | (2) | (3) | (4) | (5) | (6) | (7) | (8) | (9) | (10) | (11) | (12) | (13) | (14) |
| Assigned opportunity ($\pi$) | 0.923*** | 0.920*** | 0.913*** | 0.892*** | 0.885*** | 0.848*** | 0.825*** | 0.796*** | 0.730*** | 0.689*** | 0.629*** | 0.619*** | 0.492*** | 0.527*** |
| | (0.013) | (0.012) | (0.014) | (0.014) | (0.016) | (0.016) | (0.019) | (0.018) | (0.022) | (0.021) | (0.024) | (0.021) | (0.025) | (0.022) |
| Prior voluntary contributions ($\rho$) | 0.004* | 0.000 | 0.003+ | 0.002 | 0.004* | 0.000 | 0.006** | -0.002 | 0.005+ | -0.004 | 0.001 | -0.003 | 0.001 | 0.000 |
| | (0.002) | (0.002) | (0.002) | (0.002) | (0.002) | (0.003) | (0.002) | (0.003) | (0.003) | (0.003) | (0.003) | (0.004) | (0.003) | (0.004) |
| Student FE | Yes | Yes | Yes | Yes | Yes | Yes | Yes | Yes | Yes | Yes | Yes | Yes | Yes | Yes |
| Section-by-session FE | Yes | Yes | Yes | Yes | Yes | Yes | Yes | Yes | Yes | Yes | Yes | Yes | Yes | Yes |
| Clustered SE | Student | Student | Student | Student | Student | Student | Student | Student | Student | Student | Student | Student | Student | Student |
| Observations | 4,787 | 4,787 | 4,787 | 4,787 | 4,787 | 4,787 | 4,787 | 4,787 | 4,787 | 4,787 | 4,787 | 4,787 | 4,787 | 4,787 |
| Students | 504 | 504 | 504 | 504 | 504 | 504 | 504 | 504 | 504 | 504 | 504 | 504 | 504 | 504 |
| $R^2$ | 0.926 | 0.925 | 0.916 | 0.899 | 0.890 | 0.857 | 0.836 | 0.809 | 0.748 | 0.712 | 0.654 | 0.649 | 0.532 | 0.569 |

**Panel B. Next session**

| Dependent variable | Voluntary contributions | | | | | | |
|---|---|---|---|---|---|---|---|
| | **4 min** | **5 min** | **6 min** | **7 min** | **8 min** | **9 min** | **10 min** |
| | (15) | (16) | (17) | (18) | (19) | (20) | (21) |
| First assigned opportunity ($\beta_1$) | -0.078* | -0.080* | -0.083* | -0.089* | -0.103* | -0.119* | -0.151* |
| | (0.034) | (0.035) | (0.036) | (0.039) | (0.044) | (0.052) | (0.066) |
| Second assigned opportunity ($\beta_2$) | 0.099* | 0.102* | 0.108* | 0.116* | 0.135* | 0.151* | 0.183* |
| | (0.047) | (0.049) | (0.052) | (0.055) | (0.065) | (0.072) | (0.087) |
| Prior voluntary contributions ($\rho$) | -0.102*** | -0.102*** | -0.102*** | -0.102*** | -0.102*** | -0.102*** | -0.102*** |
| | (0.016) | (0.016) | (0.016) | (0.016) | (0.016) | (0.016) | (0.016) |
| Student FE | Yes | Yes | Yes | Yes | Yes | Yes | Yes |
| Section-by-session FE | Yes | Yes | Yes | Yes | Yes | Yes | Yes |
| Clustered SE | Student | Student | Student | Student | Student | Student | Student |
| Observations | 4,787 | 4,787 | 4,787 | 4,787 | 4,787 | 4,787 | 4,787 |
| Students | 504 | 504 | 504 | 504 | 504 | 504 | 504 |

**Panel C. All future sessions**

| Dependent variable | Voluntary contributions | | | | | | |
|---|---|---|---|---|---|---|---|
| | **4 min** | **5 min** | **6 min** | **7 min** | **8 min** | **9 min** | **10 min** |
| | (22) | (23) | (24) | (25) | (26) | (27) | (28) |
| First assigned opportunity ($\beta_1$) | -0.029 | -0.031 | -0.034 | -0.041 | -0.043 | -0.049 | -0.059 |
| | (0.049) | (0.050) | (0.051) | (0.056) | (0.063) | (0.070) | (0.087) |
| Second assigned opportunity ($\beta_2$) | 0.122* | 0.127* | 0.134* | 0.144* | 0.169* | 0.189* | 0.227* |
| | (0.050) | (0.052) | (0.055) | (0.059) | (0.070) | (0.079) | (0.096) |
| Prior voluntary contributions ($\rho$) | -0.103*** | -0.103*** | -0.103*** | -0.103*** | -0.104*** | -0.104*** | -0.103*** |
| | (0.016) | (0.016) | (0.016) | (0.016) | (0.016) | (0.016) | (0.016) |
| Student FE | Yes | Yes | Yes | Yes | Yes | Yes | Yes |
| Section-by-session FE | Yes | Yes | Yes | Yes | Yes | Yes | Yes |
| Clustered SE | Student | Student | Student | Student | Student | Student | Student |
| Observations | 4,787 | 4,787 | 4,787 | 4,787 | 4,787 | 4,787 | 4,787 |
| Students | 504 | 504 | 504 | 504 | 504 | 504 | 504 |

*Notes:* Each column redefines qualifying use as a conversation of at least the indicated number of minutes. Panel A reports the first stage, with a separate regression for each opportunity; $\pi$ is the coefficient on the assigned opportunity. In Panels B and C, the first and second assigned opportunity rows are coefficients ($\beta_1$ and $\beta_2$) from the same regression, and in Panel C the second assigned opportunity coefficient is the increment beyond the first. Figure 4 plots the second assigned opportunity rows. The ten minute columns correspond to Models 5 and 7 of Supplementary Table S4. All models include student fixed effects, section-by-session fixed effects and prior voluntary contributions, with 4,787 student-session observations from 504 students and standard errors clustered by student. Significance levels: +p < 0.1, *p < 0.05, **p < 0.01, ***p < 0.001.

**Table S6.** Cumulative effect of a second assigned opportunity on voluntary contributions, sessions 5 to 12

| Dependent variable | Voluntary contributions | |
|---|---|---|
| | ITT | LATE |
| | (1) | (2) |
| Cumulative contrast of second assigned opportunity (Δ) | 0.652*<br>(0.317) | 1.333*<br>(0.664) |
| Student FE | Yes | Yes |
| Section-by-session FE | Yes | Yes |
| Prior voluntary contributions | No | No |
| Clustered SE | Student | Student |
| Observations | 5,291 | 5,291 |
| Students | 504 | 504 |
| 95% confidence interval | [0.030, 1.274] | [0.031, 2.635] |

*Notes:* Models 1 and 2. Δ is the cumulative contrast between assignment in sessions 3 and 5 and a hypothetical assignment in session 3 alone, summed over sessions 5 through 12. The specification includes student and section-by-session fixed effects and clusters standard errors by student. It omits the prior-participation control used in Supplementary Table S4, so it uses the full 5,291 student-session two-exposure sample rather than 4,787. The 95% confidence interval equals the cumulative estimate plus or minus 1.96 times its standard error, and significance markers use the same normal approximation. Session-level differences are in Supplementary Table S7 and Figure 3. Because every student received two assignments, the one-opportunity schedule is a model counterfactual. Standard errors are in parentheses. Significance levels: +p < 0.1, *p < 0.05, **p < 0.01, ***p < 0.001.

**Table S7.** Model-predicted participation by assignment schedule (source data for Figure 3)

| Class session | ITT one-opportunity level | ITT difference (contributions) | ITT difference (%) | ITT 95% CI | LATE one-opportunity level | LATE difference (contributions) | LATE difference (%) | LATE 95% CI |
|---|---|---|---|---|---|---|---|---|
| 2 | 0.907 | 0.000 | 0.00 | [0.00, 0.00] | 0.907 | 0.000 | 0.00 | [0.00, 0.00] |
| 3 | 0.781 | 0.000 | 0.00 | [0.00, 0.00] | 0.750 | 0.000 | 0.00 | [0.00, 0.00] |
| 4 | 0.850 | 0.000 | 0.00 | [0.00, 0.00] | 0.821 | 0.000 | 0.00 | [0.00, 0.00] |
| 5 | 0.882 | 0.032 | 3.62 | [-5.48, 12.71] | 0.853 | 0.059 | 6.94 | [-10.97, 24.85] |
| 6 | 0.401 | 0.089 | 22.09 | [1.64, 42.54] | 0.371 | 0.182 | 48.98 | [2.37, 95.60] |
| 7 | 0.521 | 0.089 | 17.00 | [1.26, 32.74] | 0.492 | 0.182 | 37.02 | [1.79, 72.25] |
| 8 | 0.546 | 0.089 | 16.21 | [1.20, 31.21] | 0.517 | 0.182 | 35.19 | [1.70, 68.69] |
| 9 | 0.678 | 0.089 | 13.07 | [0.97, 25.17] | 0.648 | 0.182 | 28.07 | [1.36, 54.78] |
| 10 | 0.947 | 0.089 | 9.36 | [0.69, 18.02] | 0.917 | 0.182 | 19.84 | [0.96, 38.72] |
| 11 | 0.556 | 0.089 | 15.93 | [1.18, 30.69] | 0.526 | 0.182 | 34.57 | [1.67, 67.47] |
| 12 | 0.684 | 0.089 | 12.94 | [0.96, 24.92] | 0.655 | 0.182 | 27.78 | [1.34, 54.23] |

*Notes:* The percentage difference equals 100 times the difference between the two-opportunity and one-opportunity predictions divided by the one-opportunity predicted level in that session, which is the quantity plotted in Figure 3. The interval is the percentage plus or minus 1.96 times its standard error (the standard error of the difference divided by the same one-opportunity level). Sessions begin at 2 because session 1 is not part of the prediction sample. Levels at session 2 are the common baseline before any assignment difference, so the difference is zero.

**Table S8.** Survey items used to measure self-reported learning experiences

| Measure | Hypothesis | Item | Rating Options |
| --- | --- | --- | --- |
| Comfort participating | H2a | So far in this course, I felt comfortable speaking up and sharing my perspective during this class discussion. | 1–7 agreement |
| Mental focus | H2b | So far in this course, I felt mentally focused and engaged with our class discussions. | 1–7 agreement |
| Motivation to prepare | H2c | So far in this course, I felt motivated to prepare for and take part in the discussion in this class session. | 1–7 agreement |
| Learning of material | H2d | So far in this course, I felt that I have gained useful insights that deepen my understanding of [the course subject matter]. | 1–7 agreement |
| Four-item index | N/A | Unweighted mean of the four items. | Mean of the non-missing items; A student can have an index with fewer than four items |

*Notes:* Ratings of 1 to 7 correspond to the following options: 1 = Strongly disagree, 2 = Disagree, 3 = Somewhat disagree, 4 = Neither agree nor disagree, 5 = Somewhat agree, 6 = Agree, 7 = Strongly agree. A blank response is treated as missing for that item, and the index averages the items that were answered.

**Table S9.** Distribution of assigned opportunities by assignment window in the survey sample

| **Assignment window** | **0 assigned** N | **1 assigned** N | **2 assigned** N | **Mean assigned opportunities** | **Mean qualifying interactions** | **Students** N |
|---|---|---|---|---|---|---|
| Before the Midterm (sessions 2–5) | 203 | 337 | 87 | 0.81 | 0.37 | 627 |
| Between the Midterm and End of term (sessions 7–12) | 87 | 337 | 203 | 1.19 | 0.66 | 627 |

*Notes:* Students with both surveys (627). Qualifying interactions use the ten minute definition. Because every student was assigned twice, the two assigned counts sum to two, so the distribution between the midterm and end of term is the complement of the distribution before the midterm.

**Table S10.** Survey ratings at midterm and end of term and within-student changes

| Outcome | Midterm | | | End-of-term | | | | | | |
|---|---|---|---|---|---|---|---|---|---|---|
| | Mean | SD | N | Mean | SD | N | Lower rating % | Same rating % | Higher rating % | Paired N |
| Comfort speaking up | 5.56 | 1.34 | 627 | 5.67 | 1.29 | 627 | 22.6 | 49.1 | 28.2 | 627 |
| Mental focus | 5.86 | 1.08 | 627 | 5.83 | 1.07 | 625 | 25.0 | 51.7 | 23.4 | 625 |
| Motivation to prepare | 5.77 | 1.20 | 626 | 5.83 | 1.13 | 627 | 24.1 | 47.1 | 28.8 | 626 |
| Learning of material | 5.82 | 1.08 | 627 | 5.97 | 0.96 | 626 | 19.0 | 50.8 | 30.2 | 626 |
| Four-item index | 5.75 | 0.94 | 627 | 5.82 | 0.88 | 627 | 34.6 | 21.4 | 44.0 | 627 |

*Notes:* 627 students returned both surveys. Means, SDs and change shares use complete pairs only (Paired N). Midterm and end-of-term N are all available responses at that wave and can therefore exceed Paired N. Change percentages are the shares of students whose end-of-term rating was lower than, the same as, or higher than their midterm rating. The index is the unweighted mean of the available items. Responses are on a 1 to 7 agreement scale.

**Table S11.** Effects of CAiSEY on self-reported learning experiences by assignment window

| Dependent variable | Survey rating, points | | | | | | | | | |
|---|---|---|---|---|---|---|---|---|---|---|
| | Comfort speaking up | | Mental focus | | Motivation to prepare | | Learning of material | | Four-item index | |
| | ITT | LATE | ITT | LATE | ITT | LATE | ITT | LATE | ITT | LATE |
| | (1) | (2) | (3) | (4) | (5) | (6) | (7) | (8) | (9) | (10) |
| Midterm ($\lambda_M$) | -0.021 | -0.055 | 0.000 | 0.003 | -0.032 | -0.075 | -0.078+ | -0.183* | -0.032 | -0.076 |
| | (0.044) | (0.098) | (0.042) | (0.094) | (0.049) | (0.109) | (0.041) | (0.091) | (0.033) | (0.073) |
| End-of-term ($\lambda_E$) | 0.076* | 0.131* | -0.020 | -0.035 | 0.024 | 0.040 | 0.073* | 0.124* | 0.039 | 0.067 |
| | (0.035) | (0.063) | (0.032) | (0.056) | (0.038) | (0.066) | (0.031) | (0.055) | (0.025) | (0.044) |
| Student FE | Yes | Yes | Yes | Yes | Yes | Yes | Yes | Yes | Yes | Yes |
| Clustered SE | Student | Student | Student | Student | Student | Student | Student | Student | Student | Student |
| Observations | 1,254 | 1,254 | 1,252 | 1,252 | 1,253 | 1,253 | 1,253 | 1,253 | 1,254 | 1,254 |
| Students | 627 | 627 | 627 | 627 | 627 | 627 | 627 | 627 | 627 | 627 |

*Notes:* Models 1 through 10. $\lambda_M$ and $\lambda_E$ are the coefficients on use before the midterm and between the midterm and end of the term in Equation 2. Within each outcome, the Midterm and End-of-term rows are coefficients from the same regression. ITT is points per assigned opportunity, and LATE is points per qualifying interaction of at least ten minutes. Assignments or uses in sessions 2 through 5 enter the midterm observation (Midterm row), and those in sessions 7 through 12 enter the end-of-term observation (End-of-term row). Outcomes and regressors are deviations from the student's two-wave mean. Standard errors are in parentheses and are clustered by student. Estimates of the separate Midterm and End-of-term coefficients depend on the specification described in Methods. Significance levels: +p < 0.1, *p < 0.05, **p < 0.01, ***p < 0.001.

**Table S12.** End-of-term effects of qualifying use on self-reported learning experiences across duration thresholds

**Panel A. Comfort speaking up**

| Dependent variable | Survey rating, points | | | | | |
|---|---|---|---|---|---|---|
| | 5 min | 6 min | 7 min | 8 min | 9 min | 10 min |
| | (1) | (2) | (3) | (4) | (5) | (6) |
| End-of-term ($\lambda_E$) | 0.081* | 0.083* | 0.089* | 0.101* | 0.111* | 0.131* |
| | (0.038) | (0.039) | (0.042) | (0.048) | (0.053) | (0.063) |
| Student FE | Yes | Yes | Yes | Yes | Yes | Yes |
| Clustered SE | Student | Student | Student | Student | Student | Student |
| Observations | 1,254 | 1,254 | 1,254 | 1,254 | 1,254 | 1,254 |
| Students | 627 | 627 | 627 | 627 | 627 | 627 |

**Panel B. Mental focus**

| Dependent variable | Survey rating, points | | | | | |
|---|---|---|---|---|---|---|
| | 5 min | 6 min | 7 min | 8 min | 9 min | 10 min |
| | (7) | (8) | (9) | (10) | (11) | (12) |
| End-of-term ($\lambda_E$) | -0.021 | -0.022 | -0.023 | -0.027 | -0.029 | -0.035 |
| | (0.034) | (0.035) | (0.038) | (0.043) | (0.047) | (0.056) |
| Student FE | Yes | Yes | Yes | Yes | Yes | Yes |
| Clustered SE | Student | Student | Student | Student | Student | Student |
| Observations | 1,252 | 1,252 | 1,252 | 1,252 | 1,252 | 1,252 |
| Students | 627 | 627 | 627 | 627 | 627 | 627 |

**Panel C. Motivation to prepare**

| Dependent variable | Survey rating, points | | | | | |
|---|---|---|---|---|---|---|
| | 5 min | 6 min | 7 min | 8 min | 9 min | 10 min |
| | (13) | (14) | (15) | (16) | (17) | (18) |
| End-of-term ($\lambda_E$) | 0.025 | 0.026 | 0.028 | 0.031 | 0.034 | 0.040 |
| | (0.040) | (0.042) | (0.045) | (0.051) | (0.056) | (0.066) |
| Student FE | Yes | Yes | Yes | Yes | Yes | Yes |
| Clustered SE | Student | Student | Student | Student | Student | Student |
| Observations | 1,253 | 1,253 | 1,253 | 1,253 | 1,253 | 1,253 |
| Students | 627 | 627 | 627 | 627 | 627 | 627 |

**Panel D. Learning of material**

| Dependent variable | Survey rating, points | | | | | |
|---|---|---|---|---|---|---|
| | 5 min | 6 min | 7 min | 8 min | 9 min | 10 min |
| | (19) | (20) | (21) | (22) | (23) | (24) |
| End-of-term ($\lambda_E$) | 0.078* | 0.080* | 0.085* | 0.095* | 0.105* | 0.124* |
| | (0.033) | (0.035) | (0.037) | (0.042) | (0.047) | (0.055) |
| Student FE | Yes | Yes | Yes | Yes | Yes | Yes |
| Clustered SE | Student | Student | Student | Student | Student | Student |
| Observations | 1,253 | 1,253 | 1,253 | 1,253 | 1,253 | 1,253 |
| Students | 627 | 627 | 627 | 627 | 627 | 627 |

**Panel E. Four-item index**

| Dependent variable | Survey rating, points | | | | | |
|---|---|---|---|---|---|---|
| | 5 min | 6 min | 7 min | 8 min | 9 min | 10 min |
| | (25) | (26) | (27) | (28) | (29) | (30) |
| End-of-term ($\lambda_E$) | 0.041 | 0.043 | 0.045 | 0.051 | 0.056 | 0.067 |
| | (0.026) | (0.027) | (0.029) | (0.034) | (0.037) | (0.044) |
| Student FE | Yes | Yes | Yes | Yes | Yes | Yes |
| Clustered SE | Student | Student | Student | Student | Student | Student |
| Observations | 1,254 | 1,254 | 1,254 | 1,254 | 1,254 | 1,254 |
| Students | 627 | 627 | 627 | 627 | 627 | 627 |

*Notes:* Thresholds are 5 to 10 minutes. Each coefficient is $\lambda_E$, the effect of one qualifying interaction between the midterm and the end of the term in survey rating points, from its own regression, which also includes the Midterm term (not shown). Assignment estimates do not depend on the threshold and are in the ITT columns of Supplementary Table S11. Figure 7 plots comfort, mental focus, motivation and learning of material, and the index is reported here only. Standard errors are in parentheses and are clustered by student. Significance levels: +p < 0.1, *p < 0.05, **p < 0.01, ***p < 0.001.

**Table S13.** End-of-term effects on self-reported learning experiences in the participation cohort

| Dependent variable | Survey rating, points | | | | | | | | | |
|---|---|---|---|---|---|---|---|---|---|---|
| | Comfort speaking up | | Mental focus | | Motivation to prepare | | Learning of material | | Four-item index | |
| | ITT | LATE | ITT | LATE | ITT | LATE | ITT | LATE | ITT | LATE |
| | (1) | (2) | (3) | (4) | (5) | (6) | (7) | (8) | (9) | (10) |
| End-of-term ($\lambda_E$) | 0.077+ | 0.135+ | -0.018 | -0.031 | 0.023 | 0.038 | 0.070+ | 0.120+ | 0.039 | 0.067 |
| | (0.043) | (0.077) | (0.037) | (0.066) | (0.045) | (0.079) | (0.037) | (0.066) | (0.029) | (0.052) |
| Student FE | Yes | Yes | Yes | Yes | Yes | Yes | Yes | Yes | Yes | Yes |
| Clustered SE | Student | Student | Student | Student | Student | Student | Student | Student | Student | Student |
| Observations | 1,006 | 1,006 | 1,006 | 1,006 | 1,006 | 1,006 | 1,006 | 1,006 | 1,006 | 1,006 |
| Students | 503 | 503 | 503 | 503 | 503 | 503 | 503 | 503 | 503 | 503 |

*Notes:* The sample is the 503 survey respondents who also appear in the participation analytic sample, which differs from the 504-student sample in Supplementary Table S4. Each column is a separate regression that includes the Midterm term as in Supplementary Table S11 (only the End-of-term coefficients, $\lambda_E$, are shown). Standard errors are in parentheses and are clustered by student. Significance levels: +p < 0.1, *p < 0.05, **p < 0.01, ***p < 0.001.

**Table S14.** Balance on pre-assignment characteristics and survey response by assignment timing

**Panel A. Pre-assignment characteristics by number of CAiSEY assignments before the Midterm (survey sample)**

| **Characteristic** | **0 first-half assignments** *N = 203 students* | | **1 first-half assignment** *N = 337 students* | | **2 first-half assignments** *N = 87 students* | | **Joint test** |
|---|---|---|---|---|---|---|---|
| | Mean | SD | Mean | SD | Mean | SD | p-value |
| Pre-course comfort participating | 5.83 | 1.21 | 5.72 | 1.32 | 5.61 | 1.31 | 0.419 |
| Years of work experience | 5.19 | 1.73 | 5.45 | 2.03 | 5.51 | 2.13 | 0.259 |

**Panel B. Returning both surveys by number of CAiSEY assignments prior to the Midterm (all randomized students)**

| **Outcome** | **0 first-half assignments** | **1 first-half assignment** | **2 first-half assignments** | **Joint test** |
|---|---|---|---|---|
| Randomized students, N | 247 | 409 | 103 | |
| Returned both surveys, % | 82.2 | 82.4 | 84.5 | 0.865 |

*Notes:* Panel A tests whether pre-assignment characteristics differ across students with 0, 1 or 2 assigned opportunities to use CAiSEY prior to the Midterm (i.e., during the first half of the course: sessions 2 through 5); student counts are those of the survey sample (Supplementary Table S9). Panel B tests, among all randomized students, whether returning both surveys differs by the same grouping.

**Table S15.** Effects of CAiSEY on voluntary class participation excluding the section with reconstructed attendance

| Dependent variable | Voluntary contributions | | | | | |
|---|---|---|---|---|---|---|
| | Same session | | Next session | | All future sessions | |
| | ITT | LATE | ITT | LATE | ITT | LATE |
| | (2) | (3) | (4) | (5) | (6) | (7) |
| First assigned opportunity ($\beta_1$) | -0.005 | -0.009 | -0.066* | -0.140* | -0.027 | -0.056 |
| | (0.037) | (0.074) | (0.032) | (0.067) | (0.048) | (0.089) |
| Second assigned opportunity ($\beta_2$) | -0.022 | -0.042 | 0.101* | 0.201* | 0.106* | 0.213* |
| | (0.034) | (0.064) | (0.044) | (0.089) | (0.048) | (0.098) |
| First plus second ($\beta_1 + \beta_2$) | | | | | 0.079 | 0.156 |
| | | | | | (0.074) | (0.143) |
| Prior voluntary contributions | -0.096*** | -0.096*** | -0.095*** | -0.096*** | -0.097*** | -0.097*** |
| | (0.016) | (0.016) | (0.016) | (0.016) | (0.016) | (0.016) |
| Student FE | Yes | Yes | Yes | Yes | Yes | Yes |
| Section-by-session FE | Yes | Yes | Yes | Yes | Yes | Yes |
| Clustered SE | Student | Student | Student | Student | Student | Student |
| Observations | 4,606 | 4,606 | 4,606 | 4,606 | 4,606 | 4,606 |
| Students | 479 | 479 | 479 | 479 | 479 | 479 |
| $R^2$ | 0.378 | 0.378 | 0.380 | 0.380 | 0.379 | 0.379 |

*Notes:* Specifications replicate Models 2 through 7 of Supplementary Table S4 after excluding the retained section in which attendance was reconstructed from recorded contributions (Methods). In the All Future Sessions models, the first plus second row reports the sum of the two coefficients with its standard error. Standard errors are in parentheses. Significance levels: +p < 0.1, *p < 0.05, **p < 0.01, ***p < 0.001.

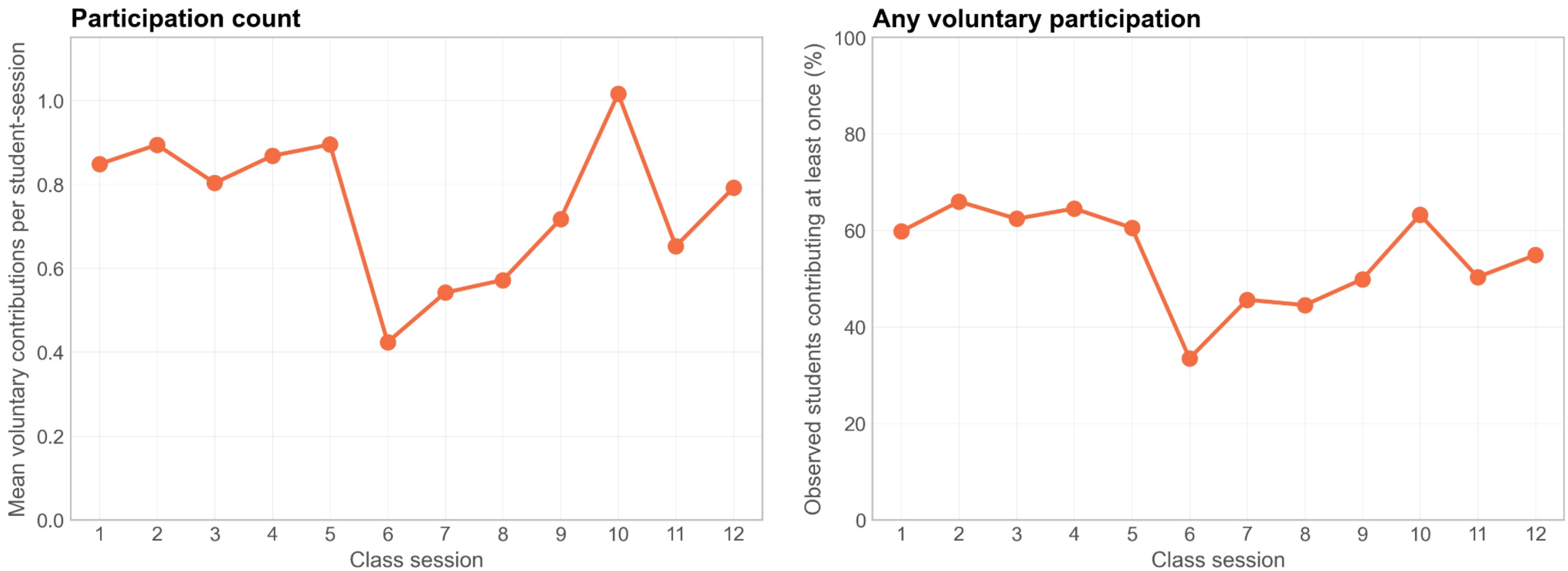

**Figure S1. Voluntary class participation across class sessions**. Left: Mean number of voluntary contributions per student-session. Right: Percentage of observed students who made at least one voluntary contribution in each session. Voluntary contributions exclude instructor-initiated cold calls. These are unadjusted summaries of the observed participation records, not estimates of the effect of CAiSEY. Session 1 was the introductory class and Session 6 involved midterm presentations, and neither included any CAiSEY assignment.